\documentclass[NJP,10pt,singlecolumn,twoside,superscriptaddress]{revtex4-1}

\usepackage{xfrac}
\usepackage{float}

\usepackage{graphicx, verbatim, mathtools, amssymb,amsmath}
\usepackage{bbold,color} 
\usepackage{tabularx}

\font\tenbsl=cmbxsl10

\def\({\left(}
\def\){\right)}

\def\a2d{a^{\dagger 2}}
\def\b2d{b^{\dagger 2}}

\def\eq#1{Eq.~(\ref{eq:#1})}

\def\fig#1{Fig.\ref{fig:#1}}

\newcommand\bra[1]{\left\langle\,#1\,\right|} 
\newcommand\ket[1]{\left|\,#1\,\right\rangle}
\newcommand\scalprod[2]{\left\langle\,#1\,\right|\left.#2\,\right\rangle}

\def\og{\leavevmode\raise.3ex\hbox{$\scriptscriptstyle\langle\!\langle$}}
\def\fg{\leavevmode\raise.3ex\hbox{$\scriptscriptstyle\,\rangle\!\rangle$}}

\def\refer#1#2{[{\tenbsl{#1}
\setbox100=\hbox{#2}\ifdim\wd100>10pt\kern .3em\box100$\,$\fi}]}

\def\lejour{le\ {\the\day}\
\ifcase\month\or janvier\or f\'evrier\or mars\or avril\or mai\or juin\or
juillet\or ao\^ut\or septembre\or octobre\or novembre\or d\'ecembre\fi\
{
\the\year}}

\def\boxit#1#2{\setbox1=\hbox{\kern#1{#2}\kern#1}%
\dimen1=\ht1 \advance\dimen1 by #1 \dimen2=\dp1 \advance\dimen2 by #1
\setbox1=\hbox{\vrule height\dimen1 depth\dimen2\box1\vrule}%
\setbox1=\vbox{\hrule\box1\hrule}%
\advance\dimen1 by .4pt \ht1=\dimen1
\advance\dimen2 by .4pt \dp1=\dimen2 \box1\relax}

\begin{document}

\author{Miller Eaton}
\email{me3nq@virginia.edu}
\affiliation{Department of Physics, University of Virginia, 382 McCormick Rd, Charlottesville, VA 22903, USA}
\author{Rajveer Nehra}

\affiliation{Department of Physics, University of Virginia, 382 McCormick Rd, Charlottesville, VA 22903, USA}
\author{Olivier Pfister}
\affiliation{Department of Physics, University of Virginia, 382 McCormick Rd, Charlottesville, VA 22903, USA}

\title{Non-Gaussian and Gottesman-Kitaev-Preskill state preparation by photon catalysis}

\begin{abstract}

Continuous-variable quantum-computing (CVQC) is the most scalable implementation of QC to date but requires non-Gaussian resources to allow exponential speedup and quantum correction, using error encoding such as Gottesman-Kitaev-Preskill (GKP) states. However, GKP state generation is still an experimental challenge.   We show theoretically that photon catalysis, the interference of coherent states with single-photon states followed by photon-number-resolved detection, is a powerful enabler for non-Gaussian quantum state engineering such as exactly displaced single-photon states and $M$-symmetric superpositions of squeezed vacuum (SSV), including squeezed cat states ($M=2$). By including photon-counting based state breeding, we demonstrate the potential to enlarge SSV states and produce GKP states. 
\end{abstract}
\maketitle
\section{Introduction}

Quantum Computing (QC) offers the  possibility to solve certain computational problems which are intractable in the realm of classical computation~\cite{Feynman1982,Shor1994}. In the last couple of decades, QC has been widely explored over discrete variables, mostly qubits, and several architectures have been proposed and experimentally realized~\cite{Ladd2010}. 
Another equally universal flavor of QC makes use of continuous variables (CV)~\cite{Lloyd1999,Braunstein2005a,Weedbrook2012,pfister2019continuous}, such as the position and momentum of a quantum harmonic oscillator or, analogously, the amplitude- and phase-quadrature amplitudes of the quantized electromagnetic field. The interest of CVQC comes primarily from the large-scale, and highly scalable, implementations that have been experimentally demonstrated of measurement-based QC substrates such as 1D cluster entangled states with at least 60 simultaneously entangled qumodes~\cite{Chen2014} and one million sequentially entangled qumodes, accessible two at a time~\cite{Yoshikawa2016}. Recently, 2D cluster states have been demonstrated~\cite{Larsen2019,Asavanant2019}. A fault tolerance threshold has been proven to exist for CVQC~\cite{Menicucci2014ft} for {the} Gottesman-Kitaev-Preskill (GKP) quantum error correction protocol~\cite{Gottesman2001,Ghose2007}. This protocol uses error code states which have a non-Gaussian Wigner function, a required resource in CV quantum information for CVQC exponential speedup~\cite{Bartlett2002}, for entanglement distillation~\cite{Eisert2002} and Bell inequality violation~\cite{Bell1987}, and for quantum error correction~\cite{Niset2009}. 

\begin{figure}[htb!]
\centering
\includegraphics[width=0.8\columnwidth]{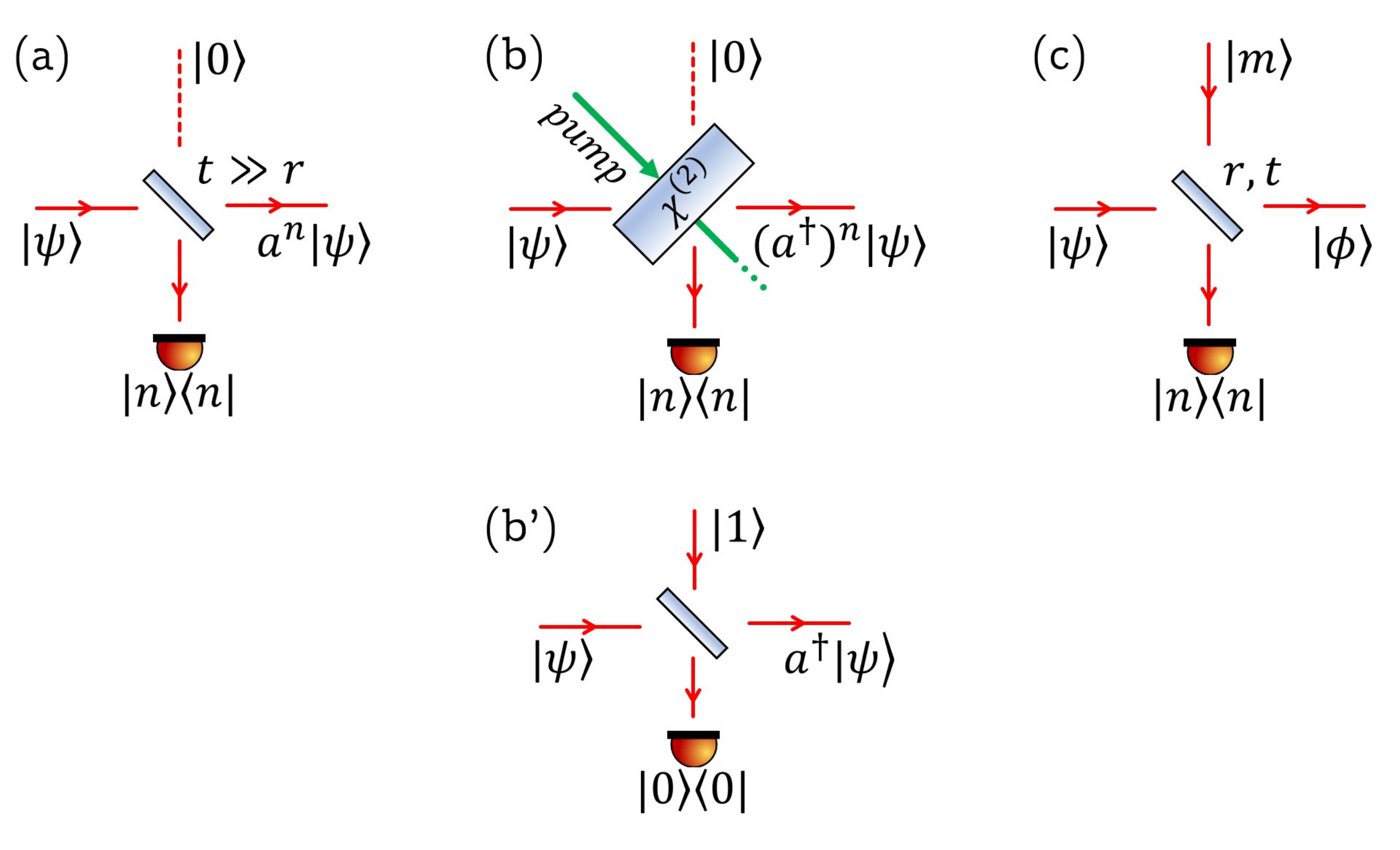}
\caption{\label{fig:phot}Various feasible techniques for generating quantum states with non-Gaussian Wigner functions. Diagram (a) depicts $n$ photon subtraction, (b) and (b') denote variants of $n$ and single-photon additions, respectively, and (c) generalizes the previous cases for an $m$ fock state with arbitrary beampslitter parameters and $n$ photon detection.}
\end{figure} 
The experimental generation of non-Gaussian states is therefore a key effort in quantum information science. It has been shown that GKP states could be probabilistically generated from squeezed Schrodinger cat states~\cite{Vasconcelos2010}, and this process has been made deterministic~\cite{weigand_generating_2018}. Single-photon states, which have been generated and characterized using heralding detection of downconverted photon pairs~\cite{Lvovsky2001,Laiho2010,Morin2012,Nehra2019}, are excellent non-Gaussian ingredients. Sophisticated techniques, such as photon subtraction~\cite{Dakna1998,Ourjoumtsev2006} and addition~\cite{Zavatta2004}, have led to very promising advances. 
In photon subtraction, \fig{phot}(a), a nonclassical state of light impinges onto a highly unbalanced, say transmissive, beamsplitter and a photon-number-resolving (PNR) detector, ideally, or at least a single-photon sensitive one, detects the reflected light. Conditioned on the detection of $n$ photons, one can show that the transmitted light is, to a good approximation, the state $a^{n}\ket\psi$, where $\psi$ denotes the initial state of light and $a$ the photon annihilation operator~\cite{Wenger2004,Ourjoumtsev2006}. This was recently generalized to multimode light~\cite{Averchenko2016}. Photon subtraction cannot work on coherent states, of course, but photon addition, Figures(b,b'), does~\cite{Zavatta2004}. In \fig{phot}(c), a more general process is presented, called photon catalysis~\cite{Lvovsky2002}.

The technique of photon catalysis is derived from the ``quantum scissors'' scheme~\cite{Pegg1998} and consists in interfering a quantum input (not necessarily pure) with a Fock state and performing PNR detection of one beamsplitter output. The beamsplitter is no longer necessarily balanced, and its coefficients are parameters of the process as well. Bartley et al. explored this process and showed its potential for the creation of non-Gaussian coherent state superpositions and squeezed states ~\cite{bartley_multiphoton_2012}.  By varying the parameters of the beamsplitter, Bartley et. al experimentally performed single-photon catalysis on a coherent state input to demonstrate the creation of states exhibiting nonclassical photon statistics.  Photon catalysis has also been explored in utilizing multi photon detection~\cite{hu_multiphoton_2016}, with general single-mode Gaussian inputs~\cite{Birrittella2018}, and for use as an entanglement enhancer~\cite{bartley_directly_2015}. Note that near-unity efficiency PNR is now an experimentally available resource, thanks to superconducting transition-edge sensors~\cite{Lita2008}.

In this paper, we explore the use of single-step and sequential photon catalysis to generate non-Gaussian states of interest to CVQC, in particular exact displaced single-photon states  and phase-symmetric superpositions of squeezed vacuum (SSV) states, of which squeezed Schr\"odinger cat states are a subset. By including a PNR-based breeding protocol, we demonstrate the potential to enlarge SSV states and generate GKP states. In this method, no squeezed states are required, the needed resources being coherent states and linear optics, single-photon states, and PNR detection.
\section{Photon catalysis}\label{sec:FSF}

In the rest of this paper, we take the restriction of photon catalysis in \fig{phot}(c) to $m$=1, i.e., to single-photon input resource states, but we keep the option of variable $n$ for PNR detection. For an arbitrary input mode $\ket\psi_{a}\ket1_{b}= \sum_{m=0}^{\infty}\psi_m|m\rangle_a |1\rangle_b$, the output state is~\cite{Caves1980,BarnettRadmore1997} 

\begin{small}

\begin{equation}
\ket{\text{out}}_{ab} = \sum_{m=0}^{\infty} \psi_m \sum_{k=0}^{m}  \binom{m}{k}^\frac12 r^{m-k}t^k \Bigl[ t\sqrt{m-k+1} |m-k+1\big \rangle_a |k\big \rangle_b 
- r\sqrt{k+1}|m-k\big \rangle_a|k+1\big \rangle_b \Bigr],
\label{eq:1}
\end{equation}
\end{small}
where the beamsplitter operator is defined by $U_{ab}= \exp[\theta(ab^\dag - a^\dag b)]$ with the reflection and transmission coefficients being $r=\cos{\theta}$ and $t=\sin{\theta}$. 
If, say, output $a$ is sent to a PNR detector which measures $n$ photons, then output $b$ is projected into the state

\begin{equation}
\begin{aligned}
\ket{\phi}_b &=\, _{a}\!\bra n out\,\rangle_{ab}\\
&=
\sum_{\ell=0}^\infty \frac{\psi_{\ell+n-1}}{\sqrt{n}}\binom{\ell+n-1}{\ell}^\frac12\,r^{n-1}t^{\ell-1}\left(nt^2-\ell r^2\right)\ket\ell_b.
 \label{eq:2}
\end{aligned}
\end{equation}

If the beamsplitter is designed so that destructive quantum interference $nt^2=\ell r^2$  occurs, then the corresponding Fock state $\ket{nt^2/r^2}$ is absent from the output. An application of this situation is given in the next section on exact state displacement by such Fock-state filtering. Note that the state amplitudes are shifted by $n-1$ in the process. When postselecting on $n=1$, this shift disappears and, by setting $r^2 = 1/(q+1)$ for the beamsplitter, one can remove the $q$-photon amplitude: 
\begin{align}
\ket{\phi_{\bar q}}_b = 
\sum_{\ell=0}^\infty \psi_\ell\,\left(\frac q{q+1}\right)^{\frac{\ell+1}{2}}\left(1-\frac\ell q\right)\ket\ell_{b}.
 \label{eq:3}
\end{align}

If $\ket\psi$ has a maximum number $\ell_\text{max}$ of amplitudes and if $q\gg\ell_\text{max}$, then the Fock-state filtering is almost perfect: only $\psi_q$ is removed and the other amplitudes are practically unchanged.

However, the number of free parameters allows us to Fock-filter a state in several different ways. Rather than postselecting $n=1$ and tuning the beamsplitter, it would be much more advantageous, from an experimental point of view, to postselect as little as possible and use PNR detection to the fullest. Let us also  fix $r=t=\frac{1}{\sqrt{2}}$, then
\begin{align}
\ket\phi_{b} \propto \sum_{\ell=0}^\infty \psi_{\ell+n-1}\binom{\ell+n-1}{\ell}^\frac12\,2^{-\frac\ell2}\left(n-\ell\right)\ket\ell_{b},
 \label{eq:4}
\end{align}
thereby removing the $n$-photon Fock state from the $b$ output if $n$ photons are detected in port $a$. Again, one should keep in mind that the state amplitudes $\psi_k$ are shifted by $n-1$ in the process, hence this operation of photon catalysis is more complex than just Fock-state filtering. However, as we now show, photon catalysis can nontrivially generate {\em exact} displaced single-photon states, as well as arbitrarily good approximations of non-Gaussian states that are of interest in quantum information processing, such as Schr\"odinger-cat states and  GKP quantum error code resource states.  In all the rest of the paper, we will consider either a coherent-state input, $\ket\psi=\ket\alpha$, or inputs derived from previous photon catalysis steps, $\ket{\psi'}=\ket{\phi}$.

To practically apply photon catalysis to arbitrary quantum states, we developed a numerical procedure to take an input density matrix, possibly impure, and transform it by photon catalysis. These calculation details are described in Appendix A.

\section{Exact displaced single-photon states}
\label{sec:DisplacedFock} 
The use of phase space displacements is very important for implementing a variety of operations in continuous-variable quantum information, since the Weyl-Heisenberg group of field quadrature shifts is the CV analog of the Pauli group for qubits~\cite{Bartlett2002}. These displacements can be experimentally realized by combining the state to be displaced with a coherent state at a highly unbalanced beamsplitter~\cite{Paris1996}. 
However, this method does require a partial trace over the other output port, yielding a statistical mixture that only approaches the exact displaced state as the reflectivity approaches unity, which in turn limits the amount of displacement. Moreover, as discussed in~\cite{kunal2018}, the displacement operation becomes less accurate as the energy constraint for the input states increases. By contrast, photon catalysis is not subject to this limitation and can generate an exact displaced single-photon state. 

\subsection{Lossless case}
We recall the derivation of Ref. \citenum{Paris1996} for an $\ket{\alpha}\ket{1}$ input.  The photon catalysis output is, before PNR detection, 

\begin{align}
   \ket{\text{out}}_{ab} = D_a(r\alpha)D_b(t\alpha)(ta^\dag-rb^\dag)\ket{0}_a\ket{0}_b.\label{eq:vac_coherent_entagle}
\end{align}
Performing a partial trace over mode $a$ results in a mixture of displaced single-photon and weakened coherent given by
\begin{equation}\label{eq:approx_disp}
    \rho_{out}= t^2\ket{t\alpha}\bra{t\alpha}+r^2D(t\alpha)\ket{1}\bra{1}D^\dag(t\alpha).
\end{equation}
In the limiting case of $\alpha \rightarrow \infty$ while $t \rightarrow 0$, we recover a pure single-photon state displaced by $t\alpha$. In all experimental cases where $\alpha$ is finite and $t>0$, we see that there is unavoidable mixing with a coherent state of the same amplitude as the desired displacement~\cite{Paris1996}.  Defining the fidelity between $\rho_{out}$ and a target state, $\rho_T$, by

\begin{align}\label{eq:fid}
    F &=\Bigg|Tr\sqrt{\sqrt{\rho_{out}}\rho_T\sqrt{\rho_{out}}}\Bigg|^2,
\end{align}

we see that the fidelity of the approximate displaced photon given with the ideal pure state, $\rho_T=D(t\alpha)\ket{1}\bra{1}D^\dag(t\alpha)$, is simply $F=r^2$.

However, if we don`t perform the partial trace and instead send mode $a$ of Eq. \ref{eq:vac_coherent_entagle} to a PNR detector and condition on the detection of $n$ photons, we get 
\begin{align}
    \ket\phi_b&\propto {}_a\!\scalprod n{\text{out}}_{ab} \\
    &\propto (nt-\alpha r^2 b^\dag)\ket{t\alpha}_b. \label{eq:nout}
\end{align}
We can then evaluate the overlap of $\ket\phi_b$ with a single-photon state displaced by $\beta$:
\begin{align}
    \bra1 D^\dag(\beta)\ket\phi & \propto nt(t\alpha-\beta)-\alpha r^2[1+\beta^*(t\alpha-\beta)].
\end{align}
Normalizing \eq{nout} and taking $\alpha,\beta\in\mathbb R$ yields the fidelity
\begin{align}
    F &= e^{-|\beta-t\alpha|^2}\frac{\big|nt(\beta-t\alpha)+r^2\alpha(1+t\alpha\beta-|\beta|^2)\big|^2}{r^4|\alpha|^2+t^2(n-r^2|\alpha|^2)^2}.
    \label{eq:8}
\end{align}
Examining this result, we see that if $r^2|\alpha|^2$ is specifically chosen to be an integer so that $n=r^2|\alpha|^2$ photons are detected, then $F=1$ for $\beta = \sqrt{|\alpha|^2 - n} = \alpha t$. While this result holds for any value of displacement, $\beta$, the specific case of taking an integer value for $|\beta|^2$ allows us to view the method of photon catalysis to enact displacements as a process that removes a single Fock component as per Eq.~\ref{eq:2}. To see this, consider the displaced single-photon state,
\begin{align}
D(\beta)\ket{1}&=D(\beta)a^\dag D^\dag(\beta)\ket{\beta}\\
&=a^\dag\ket{\beta}-\beta^*\ket{\beta}\\
&\propto\sum_{m=1}^\infty\frac{\beta^m}{\sqrt{m!}}\bigl(m-|\beta|^2\bigr)\ket{m}-|\beta|^2\ket{0}.
\end{align}

Clearly, the $m=|\beta|^2$ Fock-component is absent, which is precisely the component that was removed from the initial coherent state, $\alpha$, by the photon catalysis process, i.e., the component $nt^2/r^2=|\alpha|^2t^2=|\beta|^2$.

Our method gives the same result as the limit case given by Ref.~\cite{Paris1996}; however, here, the beamsplitter retains $t>0$ and the limit case is not needed to reach exact displacement. Therefore, we see that by tuning the reflectivity of a beamsplitter and post-selecting on the desired $n$ detection, it is possible to use photon catalysis with a coherent state to prepare an exact displaced single-photon state of  displacement amplitude  $\beta=t\alpha$. 

\subsection{Lossy case}

If we now consider an imperfect detector of quantum efficiency $\eta<1$, the final state given by \eq{rho_loss} is no longer pure and the fidelity with the target displaced Fock-state is no longer unity. However, the fidelity can be improved slightly by modifying the displacement of the target state, $\rho_T=D(\beta)\ket{1}\bra{1}D(\beta)^\dag$.  In this case, we postulate, and numerically verify, that the maximum fidelity is achieved when 
\begin{align}
    \alpha&=\sqrt{\frac{n}{\eta r^2}} \\
    \beta &= \sqrt{\alpha^2 - \frac{n}{\eta}}
\end{align}
and this value approaches unity as $t \rightarrow 0$ ($r\rightarrow1$) as illustrated in \fig2(c), 
\begin{figure}
    \centering
    \includegraphics[width=0.55\textwidth]{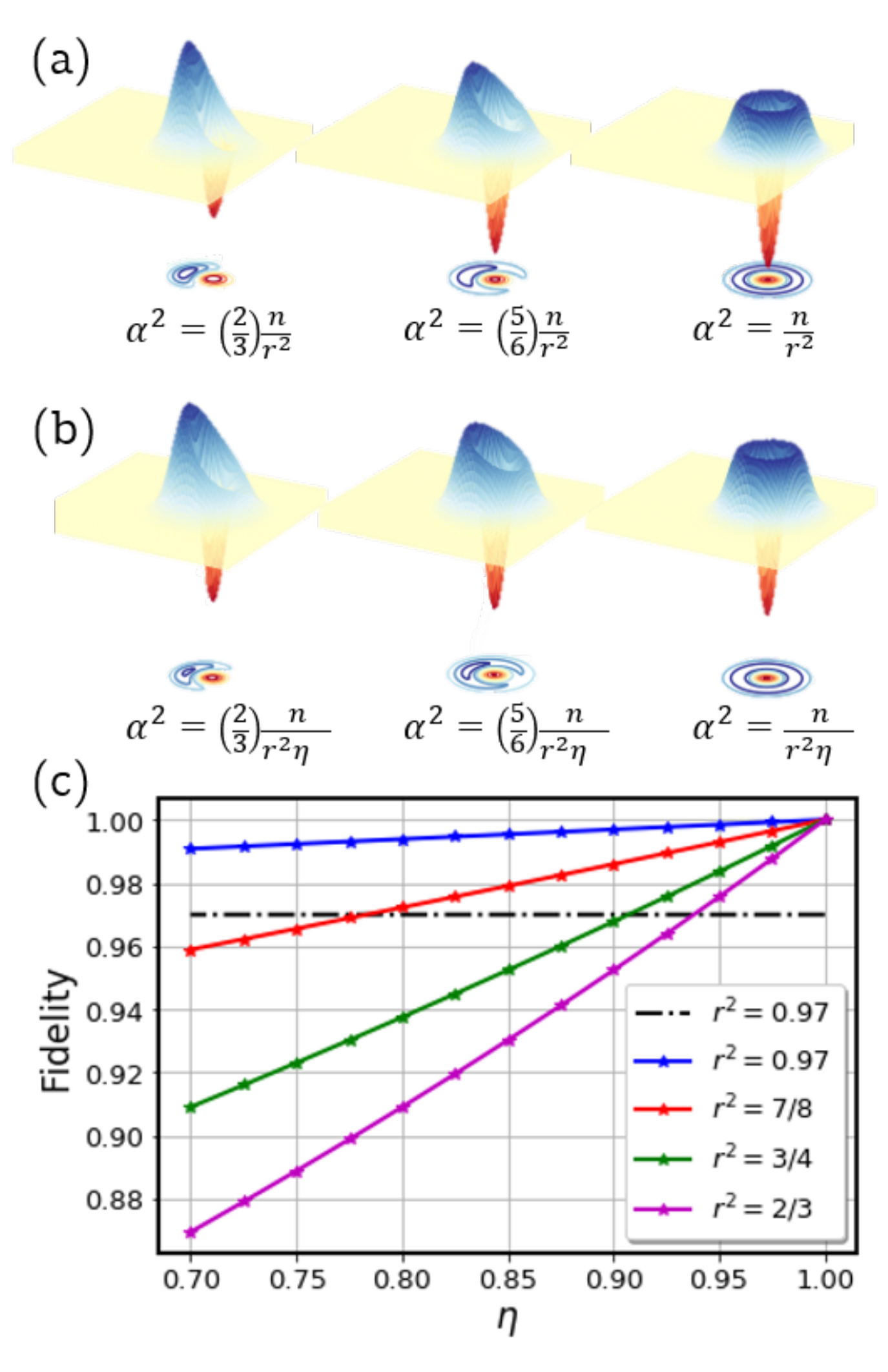}
    \caption{(a): Wigner functions for lossless photon catalysis per \eq{nout}. As the amplitude of the initial coherent state approaches the optimal ratio, the output Wigner function becomes that of a displaced single-photon Fock State.  When $\alpha^2=n/r^2$, the fidelity between the two states is unity. (b): Wigner functions for  photon catalysis with detector efficiency $\eta=0.9$. (Amplitudes have the same ratio of optimal values as in (a))  (c): Maximum fidelity achievable with a displaced Fock-state of displacement $\beta=\sqrt{\alpha^2 - n/\eta}$ for detector efficiency $\eta$ when the beamsplitter parameter $r$ is varied. The black dot-dashed line indicates the fidelity of an ideal displacement with that obtained by the usual technique~\cite{Paris1996} for $r^2=0.97$, where the PNR detector is replaced by a partial trace.}
  \label{fig:2} 
\end{figure}
where the Wigner functions for various parameters are evaluated numerically  using the open-source Python modules QuTip~\cite{Johansson2012} and Strawberry Fields~\cite{Killoran2019}.  In order to verify that the codes perform as expected, we computed the above photon catalysis step using QuTip and confirmed that the output state is identically a displaced Fock state for correctly chosen parameters.  


\fig2 shows the resulting Wigner functions after photon catalysis as the coherent state amplitude is tuned.  If the PNR detector is perfect (\fig2(a)), then the negativity of the Wigner function increases and becomes maximal as $|\alpha|^2$ approaches the optimal value to achieve an ideal displacement.  \fig2(b) demonstrates that when $|\alpha|^2$ is the same fraction of the optimal value as in \fig2(a), but the detector is no longer perfect ($\eta=0.9$), the Wigner function has the same qualitative shape in each case but exhibits an overall decreased negativity due to the effective loss at the detector. When comparing to a typical experimental displacement such as using the beamsplitter parameter $r^2=0.97$~\cite{Nehra2019}(black dot-dashed line in \fig2(c)), the addition of even imperfect PNR detectors improves the attainable fidelity with the ideal displaced single-photon state.

\section{Schr\"odinger cat states}\label{sec:CatState}

We now turn to Schr\"odinger-cat coherent superpositions (SCSs), which are of the type
\begin{align}
    \ket{SCS_{\pm}(\zeta)} = N(\ket\zeta\pm\ket{-\zeta}),
\end{align}
where $N$ is the normalization constant. These non-Gaussian states have been proposed for quantum computing~\cite{Ralph2003} and small-amplitude optical SCSs have been created by several methods~\cite{Ourjoumtsev2006,Ourjoumtsev2007,Neergaard-Nielsen2006,Takahashi2008,Gerrits2010, etesse_experimental_2015}, but there has yet to be a reliable approach to generate larger photon-number SCSs.  Methods have been proposed to ``breed'' SCSs using two smaller SCSs, a beamsplitter, and conditional homodyne detection to create a larger SCS of amplitude $\sqrt{2}\zeta$~\cite{Lund2004,Oh2018}, including an approach of particular interest involving an iterative process similar to photon catalysis where the PNR detectors are replaced with homodyne measurements~\cite{etesse_proposal_2014}. These approaches will also require quantum memories. To our best knowledge, the largest created optical SCS to-date made use of two squeezed vacuum resources and a breeding step to create a squeezed SCS with $\zeta = 2.15$ and a fidelity of $0.74$ (according to the fidelity defined by Eq. \ref{eq:fid})~\cite{Sychev2017}.

As exemplified by the work of Sychev et al.~\cite{Sychev2017}, most SCS preparation methods result in a squeezed cat state, $S(r)\ket{SCS_{\pm}(\zeta)}$, where $S(r) = \exp[r(a^{\dag\,2}-a^2)/2]$ is the single-mode squeezing operator. Without loss of generality, we take the squeezing parameter $r\in \mathbb{R}$ for the numerical simulations in this work.

Using a procedure of iterated photon catalysis steps with only a coherent state and single photons as inputs, we show that one can create states that approach exact squeezed SCS, where the final amplitude and squeezing of the state increases with the number of photon catalysis steps.  We numerically demonstrate that for low photon numbers, the final SCS amplitude after photon catalysis increases more rapidly with each additional step than with the number of steps required by the homodyne detection based breeding protocol~\cite{Lund2004}.

The general idea to generate squeezed SCS states from multiple photon catalysis steps is somewhat counterintuitive: consider, for example, the ``odd'' SCS,
\begin{align}
|SCS_{-}(\zeta)\rangle=\frac{1}{\sqrt{\sinh{|\zeta|^2}}}\sum_{n=0}^{\infty}\frac{\zeta^{2n+1}}{\sqrt{(2n+1)!}}\ket{2n+1},
\end{align}
which only contains odd photon numbers. Na\"ively, it would be tempting to consider using photon catalysis to filter the even Fock components from a coherent state in an attempt to approximate an odd SCS. However, this approach would require cascaded  stages which would ``undo'' one another in general because of the shift of the probability amplitudes by $n-1$, at each stage, when $n>1$. Thus, the previously filtered Fock amplitudes in a cascaded scheme do reappear, in general, after the next stage. Nonetheless, and remarkably, cascaded photon catalysis {\em can} be used to generate excellent approximations to {\em squeezed} SCSs --- which are useful to generate the GKP resource states crucial to CV quantum error correction, as we will see in the next section.

\begin{figure}[hb]
    \centering
    \includegraphics[width=.8\textwidth]{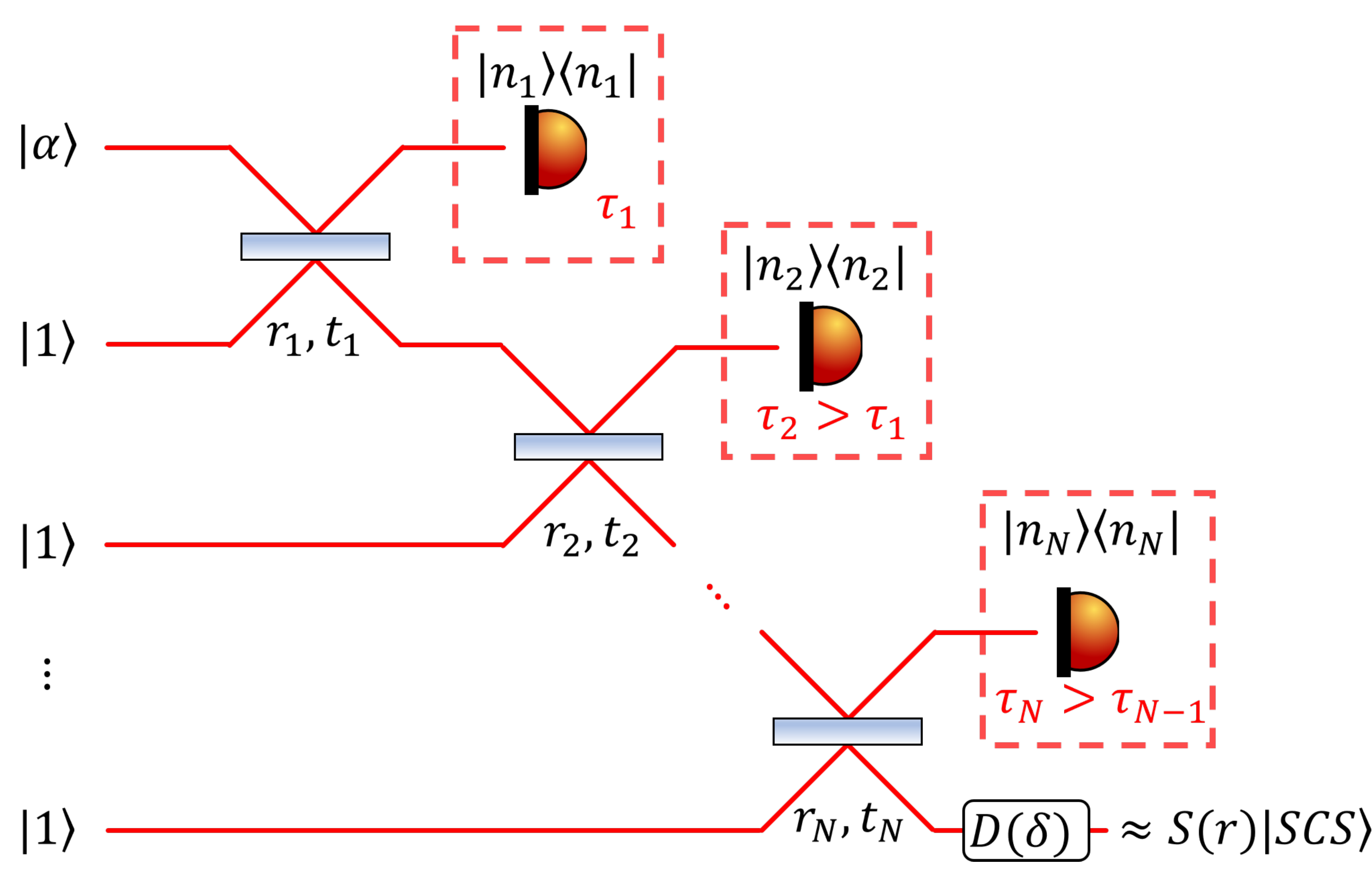}
    \caption{Protocol for generating an approximate squeezed SCS using photon catalysis. The initial coherent state of amplitude $\alpha$ is successively interfered with single-photon Fock states on a tree of variable beamsplitters, each of which is followed by a PNR detector.  These steps are sequential in that each prior PNR detection occurs before the intermediate state is interfered at the next beamsplitter in the tree.  By tuning the beamsplitter parameters and initial coherent state amplitude, the output conditioned by $N$ PNR detection steps is very near a squeezed SCS, up to a final displacement.}
  \label{fig:multiple_FSF} 
\end{figure}

\subsection{Cascaded photon catalysis}
 We turn to the squeezed SCS generation protocol, which consists of photon catalysis iterated over $N$ steps as shown in \fig{multiple_FSF}.  If we consider allowing the beamsplitter parameters and PNR measurements to vary at each step, then each iteration gives us two additional degrees of freedom: the deterministically controllable beamsplitter reflectivity, $r_i$, and the probabilistic detection, $n_i$. We limit the scope of this paper to measurements that are subsequently performed, which means that after being post-selected for an outcome $n_i$ at the $i^{th}$ step,  the output state will be the input to the $(i+1)^{th}$ step along with the single-photon as displayed in Fig.\ref{fig:multiple_FSF}. The possibility of using measurement results as feed-forward parameters, so as to optimize protocol efficiency, will be investigated in future work.

Using the general output given by Eq. \ref{eq:N_cascaded} for a coherent state, $\ket{\alpha}$, we arrive at the $N$-catalyzed state given by
\begin{equation}
\begin{aligned}\label{eq:cascaded_coherent}
&\ket{\phi{_N}} \propto
\sum_{m=0}^\infty \frac{\alpha^m}{\sqrt{m!}}\prod_{k=0}^{N-1}t^m_{k+1}\Biggl\{n_{N-k}t_{N-k}^2-r_{N-k}^2\Bigl[m+\sum_{j=0}^{k-1}(n_{N-j}-1)\Bigr]\Biggr\}\ket m. 
\end{aligned}
\end{equation}
Following the intuition that filtering Fock components may lead to an SCS, we wish to compare the fidelity of the state given by Eq. \ref{eq:cascaded_coherent} with that of a squeezed SCS.  However, in order to simplify the resulting expression and decouple the squeezing parameter from the SCS amplitude, we instead choose to compare the photon-catalyzed state with a superposition of squeezed vacuum (SSV) given by
\begin{equation}\label{eq:ssv}
    \ket{SSV_{\pm}(\beta)}\propto\left[D(\beta)\pm D(-\beta)\right]S(r)\ket0.
\end{equation}
Using the property~\cite{Walls1994} 
\begin{align}
D(\beta)S(r)&=S(r)D(\zeta)\\
\zeta&=\beta\cosh{r}+\beta^*\sinh{r},
\end{align}
it is easy to see that the squeezed SCS and SSV states are equivalent when
\begin{equation}
    \zeta=e^{-r}\beta
\end{equation}

where $\beta$ is chosen to be real without loss of generality. By allowing for a potential displacement following the multiple photon catalysis procedure, we can derive an expression for the fidelity with the $N$ step catalyzed state and the target state,
\begin{equation}
    \rho_T=D(\delta)\ket{SSV}\bra{SSV}{D^\dag(\delta)}.
\end{equation}

Here, $\delta$ is the amplitude by which the final catalyzed state must be displaced back to the origin of phase-space in order to recover the desired $|SSV\rangle$ state defined in Eq. \ref{eq:ssv}. The analytical expression for the fidelity with the target state is derived in Appendix \ref{sec:AppC} and given by
\begin{equation}
\begin{aligned}\label{eq:SSV_fidelity}
F&=\frac{C_{SSV\pm}^2C_{\ket{\phi_N}}^2}{\cosh{r}}\Biggl|\sum_{\ell,m=0}^\infty\Biggl\{(\tanh{r})^\ell\frac{(2\ell-1)!!\alpha^m}{\sqrt{(2\ell)!m!}}
[A(2\ell,m,-\beta-\delta)-A(2\ell,m,\beta-\delta))]\\
&\times\prod_{k=0}^{N-1}\Bigl[t_{k+1}^m\Bigl(n_{N-k}t_{N-k}^2-r_{N-k}^2(m+\sum_{j=0}^{k-1}(n_{N-j}-1)\Bigr)\Bigr]\Biggr\}\Biggr|^2,
\end{aligned}
\end{equation}
where $C_{SSV\pm}$ normalizes Eq. \ref{eq:ssv}, $C_{\ket{\phi_N}}$ normalizes Eq. \ref{eq:cascaded_coherent}, and we define $A(n,m,\gamma)\equiv\bra{n}D(\gamma)\ket{m}$. This expression can be optimized with respect to all available parameters. In order to limit the computation time and adhere to realistic PNR detection capabilities, we constrain each $n_i<10$. For each combination of possible detections subject to the constraints, we implemented the Nelder-Mead algorithm to perform a multivariate numerical optimization in Mathematica. Additionally, although all optimizations were performed with respect to the SSV amplitude, $\beta$, we will give the results in terms of an ideal squeezed SCS amplitude for easier comparison with other work.

\begin{figure*}
    \centering
    \includegraphics[width=\textwidth]{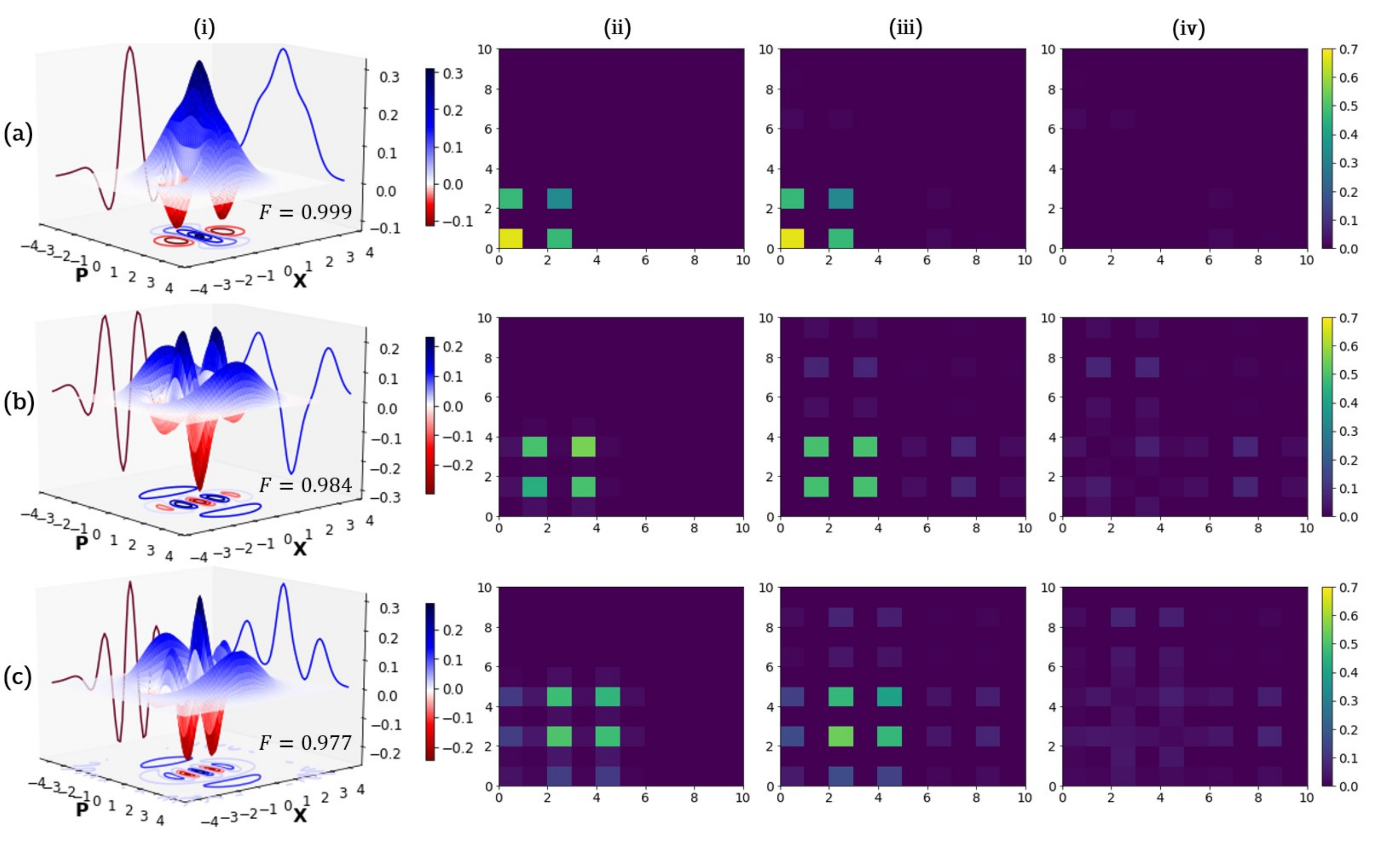}
    \caption{Calculated Wigner functions and density matrix elements for the approximate squeezed cat states resulting from varying the number of cascaded photon catalysis iterations with an initial coherent state input.  Row (a) shows the results after two iterations, (b) after three iterations, and (c) after four. In each separate case (a-c), all beamsplitter parameters, PNR detections, and the input coherent state amplitude are tuned independently to optimize the fidelity with the nearest squeezed cat state.  Column (i) shows the plotted Wigner functions for the photon catalyzed states, and (ii) gives the density matrix elements.  Column (iii) gives the density matrix elements of the ideal associated squeezed cat, and (iv) shows the magnitude of the difference between (ii) and (iii).}
  \label{fig:cat} 
\end{figure*}

We performed the fidelity optimization for a given coherent state passing through two, three, and four-step cascaded photon catalysis process. We noticed that in each case, there are multiple sets of parameters that generate a high-fidelity SSV state.  In hindsight, this may not be a surprise, as the number of free parameters likely under-constrains the system for low amplitude target states.  We selected the results that have the highest probability of success, which are depicted both in terms of the plotted Wigner function and density matrix elements in Fig.\ref{fig:cat}. We find that we can produce a nearly ideal SCS with $\zeta=1.50$ squeezed by 1.91 dB after a two step process ($F>0.999$). Optimizing a three-step process yields an SCS with $\zeta=2.18$ squeezed by 4.17 dB ($F=0.984$), and using $N=4$ results in $\zeta=2.67$ with squeezing of 4.52 dB ($F=0.977$). The full list of optimized parameters for each case is provided in Appendix~\ref{sec:data}. We note that scaling from $N=2$ to $N=3$ in this procedure already increases the cat amplitude by a factor of $\sqrt{2}$. This approach only includes a single additional interference and detection stage whereas the other breeding procedure discussed earlier requires three additional steps provided that the two smaller cats to be bred came from an $N=2$ photon catalysis procedure. This indicates that an iterative photon-catalysis process might scale better for larger SCS creation than homodyne-based breeding.

\begin{figure}
    \centering
    \includegraphics[width=0.95\textwidth]{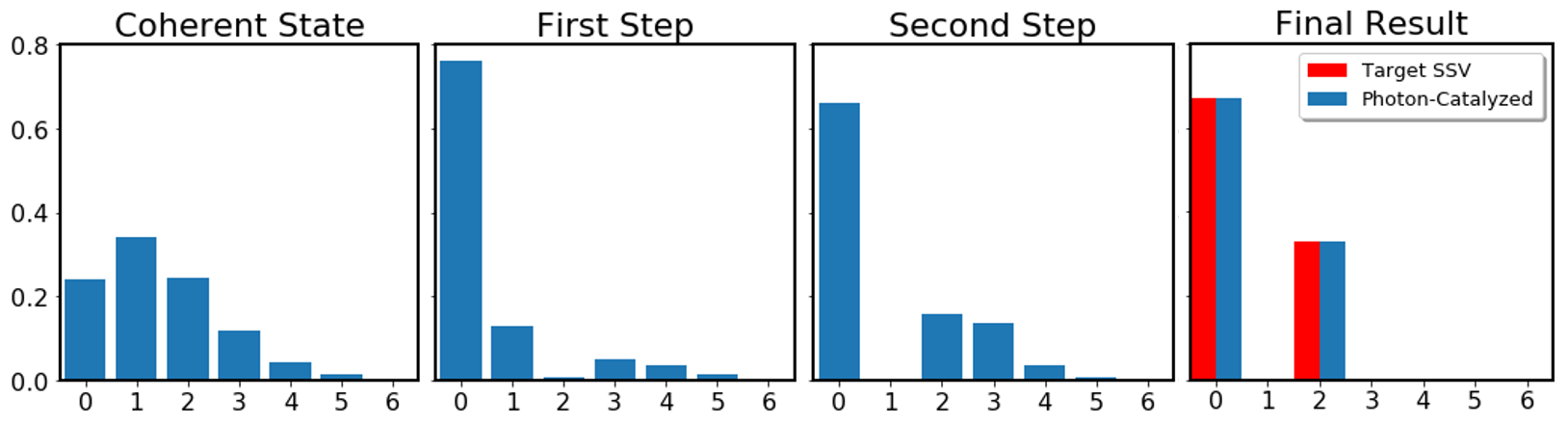}
    \caption{Photon number probability amplitudes for various states.  Top:   distributions for coherent state input and the quantum state after the first detection step. Bottom:  distributions for the quantum state after the second PNR detection and for the final catalyzed state after a displacement(blue) compared to the target SSV state(red).}
 \label{fig:ampl} 
\end{figure}

It is important to keep in mind that each $N$-step photon catalysis process is optimized to produce an approximate SSV state as a final result; simply adding an additional catalysis step to any of these final approximate cats will not increase the cat amplitude or squeezing under the current approach. As an example, consider \fig{ampl}, which shows the photon number distributions before and after each catalysis step for $N=2$.  The filtering of Fock components at each step is not perfect due to the numerical procedure, but it is easy to see that the first projective PNR measurement, $|1\rangle \langle 1|$,  acts to nearly eliminate the $\ket{2}$ component from the initial coherent state, while the second step effectively shifts and re-scales the distribution. A final displacement of the Wigner function back to the origin of phase-space results in a very near approximation to the target state as shown by the comparison of probability amplitudes in the rightmost panel of \fig{ampl}.  Although the first catalysis step acts to de-Gaussify the coherent state, applying the displacement at this point would result in poor overlap with any class of SSV.

\subsection{Loss tolerance}
\begin{figure}[!hbt]
    \centering
    \includegraphics[width=0.9\columnwidth]{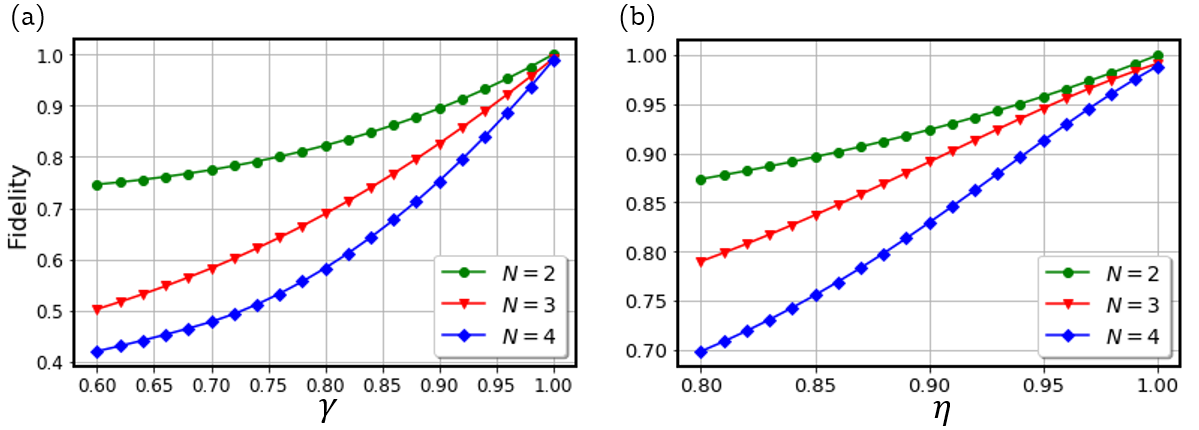}
    \caption{The effects of vacuum mixing due to (a) impure single-photon input and (b) imperfect detectors are shown separately for $N$-step photon catalysis with $N=2,3$ and $4$. The fidelity is calculated with respect to the same target SSV states as for the optimized procedure shown in Fig.\ref{fig:cat}. }
    \label{fig:mixture_input}
\end{figure}

Experimental imperfections and associated optical losses will contribute to non-ideal single-photon resources, in which case we must consider a mixture of single-photon and vacuum as the realistic input at each interference beamsplitter~\cite{Lvovsky2001, Nehra2019}. This can be modeled by sending each pure single photon through a lossy channel with transmission $\gamma$, so that each input is now
\begin{equation}
    \rho_{mix}=\gamma \ket{1}\bra{1}+(1-\gamma)\ket{0}\bra{0}.
\end{equation}
 By using the optimized parameters for the ideal procedure and replacing the pure single-photons with $\rho_{mix}$ at each step, we see in Fig.\ref{fig:mixture_input} how including a vacuum contribution adversely affects the fidelity with the target states in each case. The decreased purity more greatly harms the procedures with larger $N$, unsurprisingly. 
 
 
 However, we note that just as in Ref.~\cite{etesse_proposal_2014}, our protocol in of itself adds little imperfection to the result, which can be seen directly in Fig.\ref{fig:mixture_input} by comparing the drop in fidelity between neighboring values of $N$.  For purities above $\gamma=0.8$, the decrease in the fidelity of our output state with the target between $N=2$ and $N=3$ step processes is nearly equivalent to the fidelity difference between $N=3$ and $N=4$ step procedures, which indicates the change in fidelity is nearly linear with the number of impure single-photon inputs.
 
 Fig.\ref{fig:mixture_input} indicates the importance of pure single photons for our protocol; fortunately, there has been much work in the area of on-demand, highly efficient single-photon emitters using solid state devices~\cite{aharonovich_solid-state_2016,loredo2016scalable, somaschi2016near}. On-demand quantum dot emitters have already demonstrated single-photon purities above $99\%$~\cite{ding_-demand_2016}, and a recent proposal making use of quantum feedback control suggests that on-chip sources with greater than $99\%$ on-demand extraction efficiency are possible with current technology~\cite{heuck_temporally_2018}.

As in Sec. \ref{sec:DisplacedFock}, we consider the effects of imperfect detectors in each step, where we assume identical detector efficiencies of $\eta\leq 1$.  Similarly to having impure single-photons, each additional imperfect PNR measurement acts to further decrease the fidelity with the target state in a way that appears linear with increasing photon catalysis steps.  We see from \fig{mixture_input} that generating high-fidelity cat states would require the use of highly efficient PNR detectors, as $\eta\leq0.85$ prohibits even $N=2$ step procedures from yielding states above a fidelity of 0.9. Thanks to recent advances, high efficiency PNR is achievable with current transition-edge sensors~\cite{Lita2008} and superconducting nanowire single-photon detector arrays~\cite{dauler_review_2014}, which have been recently reported to have PNR capabilities~\cite{Nicolich2019}.

\subsection{Success probability}

For methods of non-Gaussian state engineering that use homodyne detection as in Ref.~\cite{etesse_proposal_2014}, one can select a confidence window of $\Delta x$ on acceptable post-selection measurement results that allow for an increase in procedure success probability at the cost of decreasing fidelity.  In our case, however, the projective operator of PNR detection lies in a discrete subspace, so we cannot allow for continuous deviations from the ideal detection scheme.  Nonetheless, we can still examine the fidelity and success probability in cases where the initial coherent state amplitude and $N$ deterministic beamsplitter parameters are held fixed at values optimized for SSV generation, but some $n_i$ deviate, necessarily in integer steps, from the ideal scenario. 

\begin{figure}[!h]
    \centering
    \includegraphics[width=0.65\columnwidth]{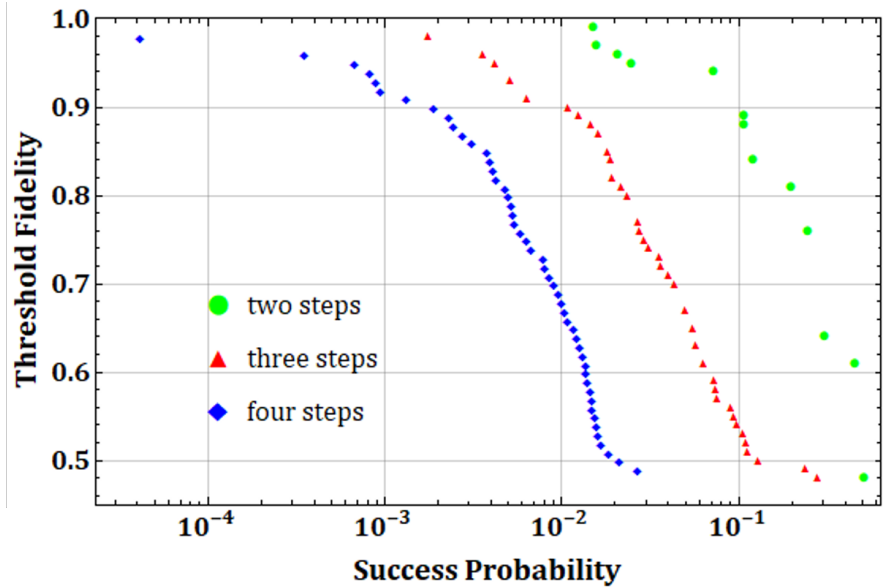}
    \caption{The success probability to generate an approximate SSV state with at least a given threshold fidelity is shown for a cascaded photon catalysis procedure of two(green), three(red), and four(blue) steps. Allowing the PNR detection results to deviate from the optimized values yields an increased probability of success at a cost to fidelity.}
    \label{fig:success}
\end{figure}

For each combination of detection events that deviates, we can re-optimize the fidelity from Eq. \ref{eq:SSV_fidelity}, where now only the target state parameters are allowed to vary (SSV amplitude, squeezing, and the displacement amplitude).  In this way, it is possible to generate a finite list of PNR detection combinations that differ from the combination that best approximates an ideal SSV state, yet still yields a final state above a given threshold fidelity, $F_{thr}$, with a target state. This suggests that approaches involving feed-forward based on measurement results might also be fruitful. If one only discards the photon-catalyzed states where the detected combination $\{n_1, n_2,\ldots, n_N\}$ is outside of the acceptable list, then the photon-catalyzed state is guaranteed to have a fidelity $F\geq F_{thr}$ with an appropriate SSV state. This target state may vary slightly in amplitude and squeezing from the SSV target for the optimal detection scheme (removing a different total photon number changes the total energy of the state), but the target is still required to be an SSV state, and we show later how two similar SSV states can be enlarged by utilizing PNR detection. By decreasing $F_{thr}$, more combinations of possible PNR detections are included in the ``accept'' list, thereby increasing the probability of successfully generating an SSV state. The effects of lowering $F_{thr}$ for the two, three, and four-step photon catalysis procedures considered in the previous section are depicted in Fig.\ref{fig:success}. For the $N=2$ and $N=3$ cases, we see that setting $F_{thr}=0.9$ yields nearly an order of magnitude increase in success probability from the perfect detection scenario, while the $N=4$ case shows an increase of nearly two orders of magnitude for the same fidelity threshold.  Paired with state-of-the-art technology in deterministic single-photon sources~\cite{ding_-demand_2016,hanschke2018quantum} with emission rates on the order of $10^6$ counts/sec~\cite{aharonovich_solid-state_2016}, an $N=3$ step photon catalysis process would yield nearly perfect SSV states at rates in the kilohertz regime.

\subsection{Higher-symmetry SSV states}
\label{sec:high_symm}
The versatility of cascaded photon catalysis allows us to examine classes of SSV states that exhibit beyond two-fold phase space rotational symmetry. These states can be more generally expressed as superpositions in the form of
\begin{equation}
    \label{eq:ssv_gen}\\
    \ket{SSV^{(M)}}\propto \sum_{n=0}^{M-1} D(\beta e^{\frac{2n}{M}i\pi})S(r e^{\frac{4n}{M}i\pi})\ket{0}, 
\end{equation}
where $M$ is the number of components in the superposition and degree of rotational symmetry. Each component consists of squeezed vacuum that has been displaced with an integer multiple of phase increment of $\sfrac{2\pi}{M}$, where the squeezer anti-squeezes the vacuum perpendicularly to the axis of the subsequent displacement. Although there is no restriction on $r$ for this class of states, we consider $r<0$, as this corresponds to the states we generated with cascaded photon catalysis. While we only explicitly consider protocols to generate states with $M=2$ and $M=3$ in this work, we expect that it is possible to further generalize the generation of $SSV^{(M>3)}$ by optimizing photon-catalysis.

\begin{figure}[h]
    \centering
    \includegraphics[width=0.8\columnwidth]{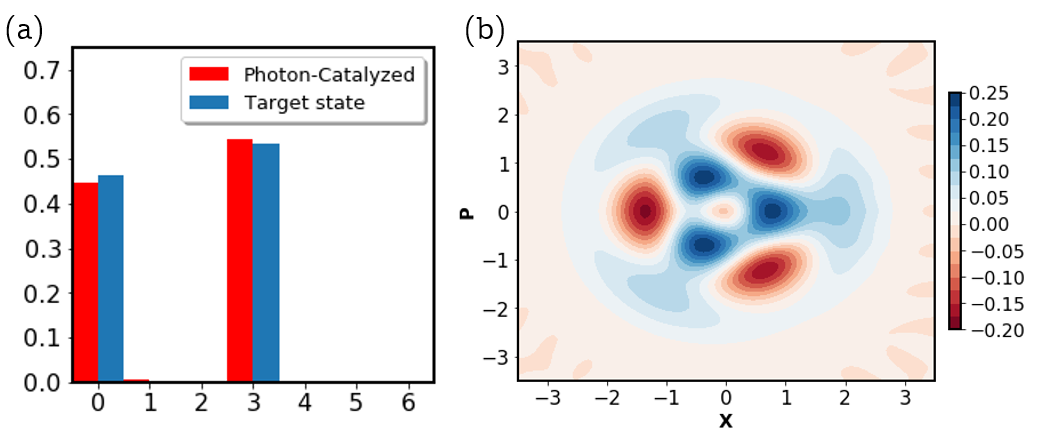}
    \caption{Photon-number distribution, (a), and Wigner function, (b), of the approximate $SSV^{(3)}$ state generated by a three-step photon catalysis protocol, where the squeezing parameter is $r=-0.24$ and amplitude is $\beta=1.255$.  The Wigner function nicely displays $120^{\circ}$ phase-space rotational symmetry.}
    \label{fig:tri_ssv}
\end{figure}

For the next highest symmetry state (three-fold symmetry), we can again use the protocol depicted in Fig.\ref{fig:multiple_FSF} with three photon catalysis steps, but with different input coherent state amplitude and beamsplitter parameters, to create the state
\begin{equation}
    \Bigl(D(\beta)S(r)+D(\beta e^{\frac{2\pi i}{3}})S(re^{\frac{4\pi i}{3}})+D(\beta e^{\frac{4\pi i}{3}})S(re^{\frac{2\pi i}{3}})\Big)\ket{0},
\end{equation}
where $r=-0.24$, $\beta=1.255$, and the fidelity is greater that $0.99$.  This is achieved with detection events of $\{n_1, n_2, n_3\}=\{6,\,4,\,2\}$, which were found serendipitously.
The comparison of the ideal and procedurally generated states is depicted in Fig.\ref{fig:tri_ssv}(a), and Fig.\ref{fig:tri_ssv}(b) demonstrates the $120^{\circ}$ rotational symmetry of the Wigner function.  As this state is created with a cascaded photon catalysis procedure, the success probability and dependence on purity of single-photon inputs and detector efficiency is very similar to the other $N=3$ case considered earlier. The utility of the $M=3$ SSV state is developed further in the next section.
 
\section{Breeding Protocols}
Methods to combine SCS with breeding protocols have been widely explored to generate larger amplitude SCS states~\cite{Sychev2017,etesse_experimental_2015,Lund2004,Oh2018,laghaout_amplification_2013} and other classes of comb states~\cite{etesse_proposal_2014,Vasconcelos2010}. A recent work shows how feed-forward displacements can allow for deterministic creation of optical GKP states from a supply of squeezed SCS~\cite{weigand_generating_2018}.  Here, we motivate the extension of these breeding protocols to PNR detection as opposed to traditional homodyne measurements.  Additionally, we make use of the phase-insensitivity of direct photon-counting measurements to demonstrate how PNR detectors can be used to breed higher symmetry states.     
\subsection{SSV states}
First, we begin by examining a procedure with the aim to take two lower amplitude M\textsuperscript{th}-order SSV states and produce a larger SSV state of the same order of symmetry. Taking two identical states given by Eq. \ref{eq:ssv_gen} and interfering them on a balanced beamsplitter, we have the state 
\begin{equation} \label{eq:breeding_input}
    \ket{\phi}\propto U_{ab}\sum_{n=0}^{M-1} D_a(\beta_n)S_a(\xi_n)\ket{0}_a\otimes\sum_{n=0}^{M-1} D_b(\beta_n)S_b(\xi_n)\ket{0}_b,
\end{equation}
where $\beta_n=|\beta| e^{\frac{2n}{M}i\pi}$ and $\xi_n=r e^{\frac{4n}{M}i\pi}$, with the beamsplitter operation given by $U_{ab}=e^{\frac\pi4(ab^\dag - a^\dag b)}$ . Developed further in Appendix \ref{sec:breeding_appen} and numerically verified for the first few SSV states, it can be seen that for finite squeezing and relatively larger values of $|\beta|$, a PNR measurement of zero photon, $|0\rangle \langle 0|$, on output mode $b$ leads to
\begin{equation}
    {}_b\langle0\ket{\phi}_{ab}\approx \sum_{n=0}^{M-1} D_a(\sqrt{2}\beta_n)S_a(\xi_n)\ket{0}_a.
\end{equation}

\begin{figure}[hb]
    \centering
    \includegraphics[width=1\textwidth]{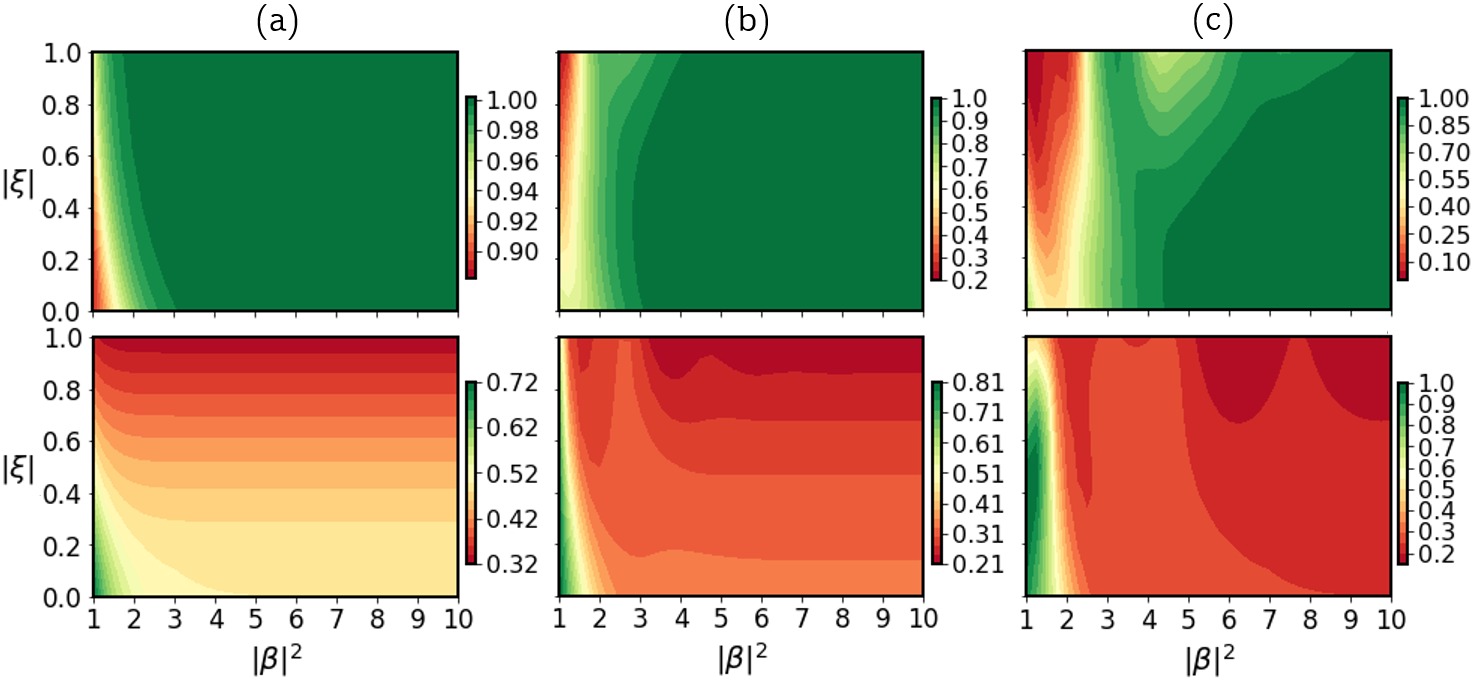}
    \caption{Fidelity and success probability of the PNR detection-based breeding protocol for various initial amplitudes, $\beta$, and squeezing, $\xi$. The colors bars indicate fidelity (top panel) and success probability (bottom panel) for the respective plots, where dark green is unity. $M=2$ (a), $M=3$ (b), and $M=4$ (c) symmetric SSV states are bred into larger states in the same class with amplitude $\sqrt{2}\beta$ and the same squeezing parameter.}
    \label{fig:breeding}
\end{figure}
This means that using PNR detectors, SSV states can be bred into states with the same squeezing, but with increased amplitude $|\beta|\rightarrow\sqrt{2}|\beta|$.  The factor of $\sqrt{2}$ increase is the same result as achieved for SSV states with $M=2$ using homodyne detection-based breeding protocols, but the nature of performing a homodyne measurment along a particular axis in phase-space means that higher symmetry features are not preserved, thus indicating that PNR-based protocols such as this will be necessary to breed general SSV-type states. The fidelity of the bred states with the enlarged target states as well as success probabilities are shown in Fig.\ref{fig:breeding}. In all cases examined, the fidelity quickly approaches unity for values of $|\beta|^2 \gtrapprox M$.  Additionally, the success probabilities are greater than $10^{-1}$ in all regions where fidelities are numerically indistinguishable from one, and the probability for success in the $M=2$ case holds steady at $0.5$ for squeezing $|\xi|\lessapprox 0.3$ and $|\beta|^2>4$.

\subsection{Square-lattice GKP states}
\begin{figure*}[ht]
    \centering
    \includegraphics[width=1\textwidth]{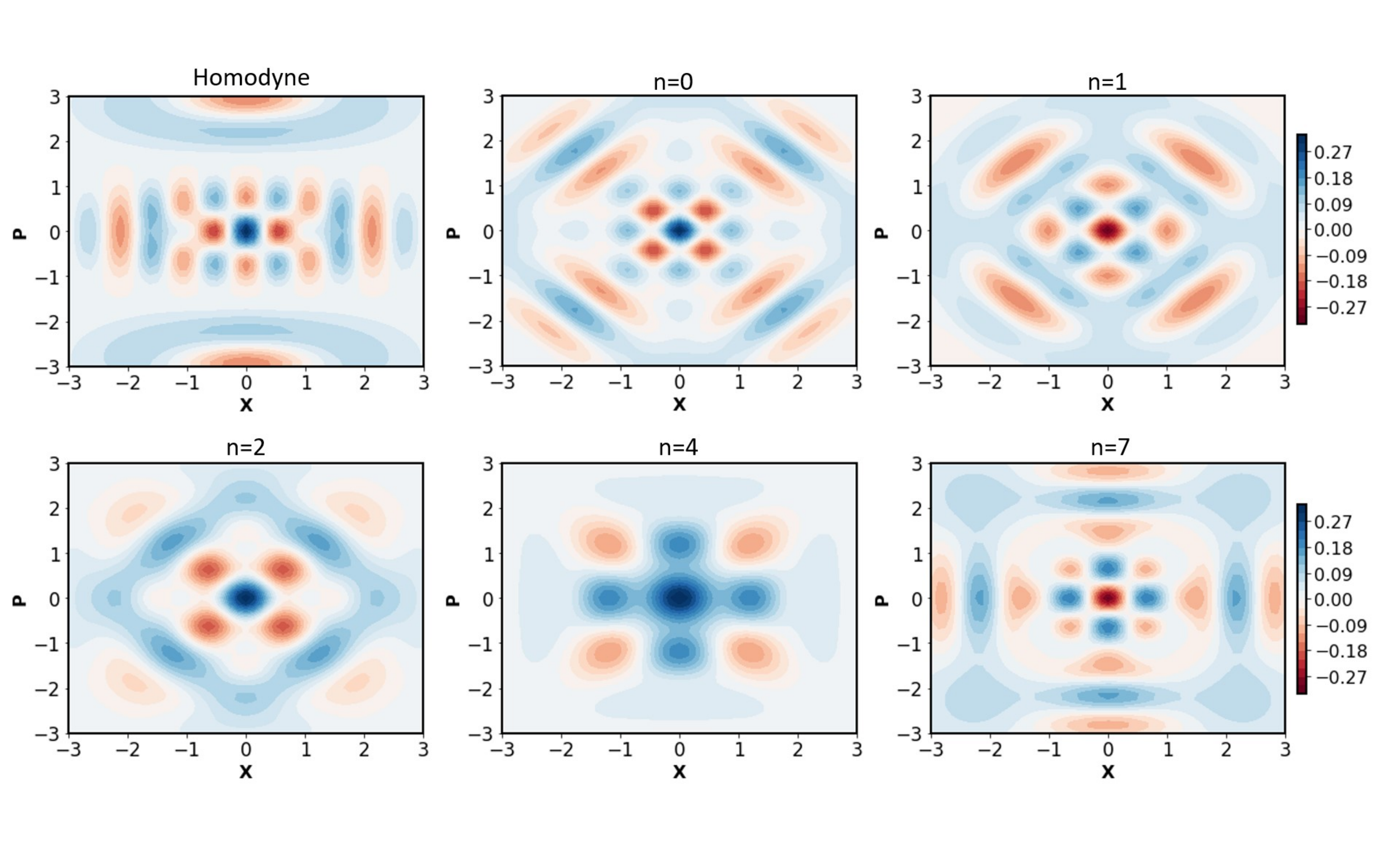}
    \caption{Wigner functions of bred two-fold symmetry SSV states (squeezed SCS states) achieved by the photon catalysis protocol after various detector measurements.  The homodyne plot demonstrates the grid state of the result after a $p=0$ quadrature measurement, while the other plots show the resultant state after a PNR detection of $n$ photons.  The $n=4$ case has a fidelity $F=0.996$ with a square-grid GKP state with finite-energy peak-width $\Delta=0.545$.}
  \label{fig:square_grid} 
\end{figure*}

Here, we adapt the experimental proposal of Vasconcelos et al.~\cite{Vasconcelos2010} to our cascaded photon catalysis method of producing SCSs followed by PNR detector-based breeding.
The essence of their original idea is as follows:  by entangling two squeezed SCS on a balanced beamsplitter and performing a homodyne measurement of $p=0$ on one output, the output of the other port is projected into a state exhibiting a series of evenly spaced Gaussian peaks along the $x$-quadrature axis where the peak amplitudes follow a binomial distribution about a central peak at $x=0$, and the peak width is determined by the SCS squeezing. Repeating an identical protocol with two of these outputs serves to increase the number of peaks by elevating the order of the binomial distribution. Recently, Weigand et. al have shown that including feed-forward displacements based on the homodyne measurement results makes this process deterministic~\cite{weigand_generating_2018}. After many iterations and in the limit of large initial SCS squeezing, this method accurately approximates an ideal square-lattice GKP state, which can be explicitly written as~\cite{Gottesman2001}
\begin{align}
    |\mu_\text{sqr}\rangle \propto \sum_{n_1,n_2=-\infty}^\infty e^{-i\hat{p}\sqrt{\frac{2\pi}{d}}(dn_1+\mu)} e^{i\hat{q}\sqrt{\frac{2\pi}{d}}n_2}|0\rangle,
\end{align}
where $d$ is the dimensionality of the code space and $\mu=0,\dots,d-1$ labels the specific GKP state in the code space. To make the state physically realizable in the case of finite energy, the state can be modulated by a Gaussian envelope~\cite{Noh2018}.  The finite energy GKP state is thus given by
\begin{align}
\ket{\mu^{\Delta}_{sqr}} \propto e^{-\Delta^2\hat{N}}\ket{\mu_{sqr}}
\end{align}

where $\hat{N} =a^\dagger a$ is the photon-number operator. These finite-energy GKP states compose a periodic square-lattice grid in phase-space, where each peak has width $\Delta$. The result of following this homodyne-detection based procedure with two states generated by our four-step photon catalysis procedure is shown in the top left plot of Fig.\ref{fig:square_grid}.   

If we perform PNR detection on one mode following the entangling beamsplitter instead of a homodyne measurement, we can obtain different final states according to the selected measurement result that each exhibit the periodic comb-like structure of a GKP state as demonstrated in Fig.\ref{fig:square_grid}. Using a balanced beamsplitter, $U_{ab}$, and phase-shift operator $R(\Theta)=e^{i\Theta a^\dag a}$, the overall result can be concisely written as
\begin{align}
    \ket{\phi}\propto {}_b\!\bra{n}U_{ab}R_a(\Theta)\ket{SSV{^{(2)}}}_a\ket{SSV^{(2)}}_b.
\end{align}
This is exactly the process that we used earlier to breed SSV states, but with two important changes.  Because PNR detection is phase-insensitive, precisely controlling the relative phases between the states impinging on the entangling beamsplitter becomes vital, and here we set the phase difference to $\Theta=\frac\pi2$ to achieve the square-grid comb-like periodicity. Additionally we allow for PNR detection other than vacuum. 

As a concrete example, if we entangle two SSV states generated from photon catalysis (Fig.\ref{fig:cat}(c)) and detect $n=4$ photons in one mode, we achieve a state that has a fidelity of $F=0.996$ with a finite-energy square GKP state of width $\Delta=0.545$. This single event occurs with a success probability of $0.09$, but the probability to achieve at least one of the comb states shown in Fig.\ref{fig:square_grid} exceeds $0.60$.

Since finite-energy GKP code states are no longer orthogonal within the $d$-dimensional code space, error correction protocols become probabilistic due to the non-zero overlap between the logically encoded states. In the original work, the probability to successfully correct an error hinges solely on the ability to accurately distinguish code states, leading to a $99\%$ success rate for $\Delta\sim0.5$~\cite{Gottesman2001}, such as is the case for our states.  However, a more realistic error-correcting protocol considers potential shift errors on the ancilla qubits as well as the encoded qubit to be corrected, leading to a more stringent requirement of $\Delta<0.15$ to achieve the same success rate~\cite{glancy2006error}.  Utilizing more photon catalysis iterations to increase the initial SCS squeezing and performing multiple breeding operations as in \cite{weigand_generating_2018} will result in suitable states for realistic quantum error correction. 

\subsection{Hexagonal GKP states}
\begin{figure*}[h!]
    \centering
    \includegraphics[width=1\textwidth]{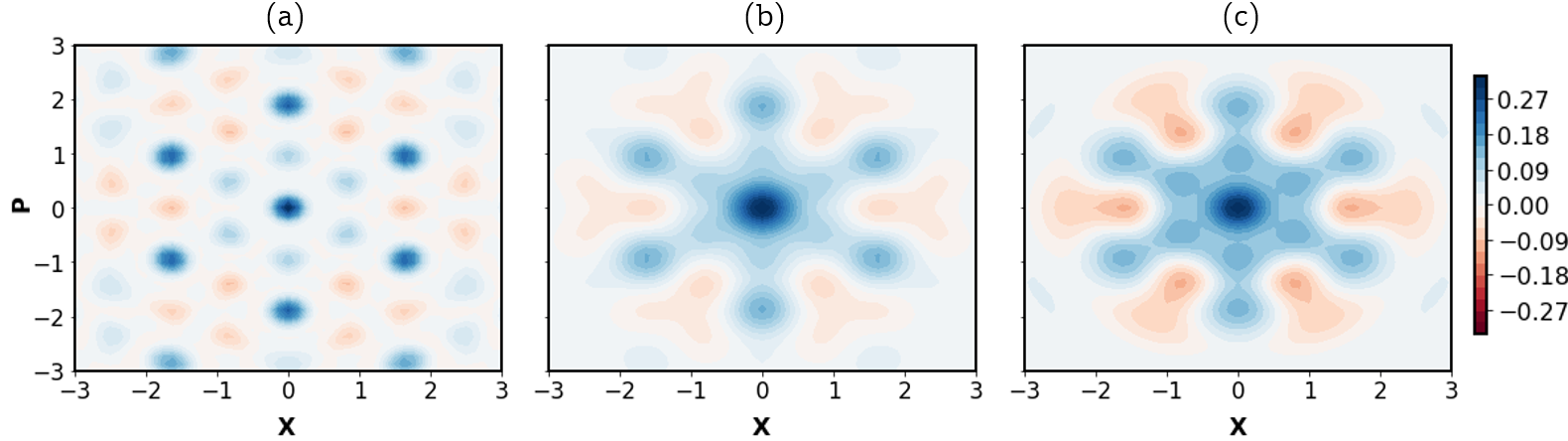}
    \caption{Comparison of finite energy hex GKP states to that of two $120^{\circ}$ rotationally symmetric SSV states after breeding. (a) and (b) show the Wigner function of a mixture of hex GKP states of code dimension 2 with peak widths $\Delta=0.2$ and $\Delta=0.46$, respectively.  (c) shows the Wigner function of bred SSV states each developed by an $N=3$ photon catalysis protocol.}
  \label{fig:hex_approx} 
\end{figure*}

One principle advantage exhibited by the GKP encoding scheme is an inherent resistance to shift errors in the conjugate position and momentum quadratures $x$ and $p$. The hex GKP state, formulated in the original work~\cite{Gottesman2001} and recently expanded upon in greater detail in Refs.~\cite{Noh2018} and ~\cite{albert_performance_2018}, improves upon the initial square lattice state by being able to correct larger displacement errors.  As a hexagonal lattice in phase space, the hex GKP state can now account for shift errors with a new upper-bound phase space radius of $r\leq\sqrt{\frac{\pi}{\sqrt{3}d}}$, as opposed to $r\leq\sqrt{\frac{\pi}{2d}}$ intrinsic to the original, where $d$ is the dimensionality of the code space. This hex GKP state can be explicitly written as
\begin{align}
    |\mu_\text{hex}\rangle \propto \sum_{n_1,n_2=-\infty}^\infty e^{-\tfrac{i}{2}(\hat{q}+\sqrt{3}\hat{p})
            \sqrt{\frac{4\pi}{d\sqrt{3}}}(dn_1+\mu)} e^{i\hat{q}\sqrt{\frac{4\pi}{d\sqrt{3}}}n_2}|0\rangle,
\end{align}
where as in the square-lattice case, applying the non-unitary operator $e^{-\Delta^2\hat{N}}$ makes the state physically realizable~\cite{Noh2018}.  Fig.\ref{fig:hex_approx} shows the Wigner function for a linear combination of hex GKP states of dimension $d=2$ plotted with different peak-widths.  

Noting that the SSV\textsuperscript{(2)} states (squeezed SCS) can be bred into square-lattice GKP states, we are motivated to examine what happens when breeding SSV\textsuperscript{(3)} states, such as those generated in Sec. \ref{sec:CatState}\ref{sec:high_symm}. Because the states have $\sfrac{2\pi}{ 3} $ rotational symmetry, we fix the phase difference of the two initial states to be $\Theta=\sfrac\pi3$ at the entangling beamsplitter, and send one output mode to a PNR detector. Just as with the SSV\textsuperscript{(2)} breeding protocol, we find that several possible measurement results lead to a final state exhibiting the desired hexagonal symmetry. Fig.\ref{fig:hex_approx}(c) shows the result of breeding two of the states depicted in Fig.\ref{fig:tri_ssv} after post-selecting on a PNR measurement of $n=0$ with a probability of 0.31. While only having a fidelity of $0.8$ with the mixed hex GKP state in Fig.\ref{fig:hex_approx}(b), the visual similarity and relatively high success probability indicate that there may be a suitable SSV\textsuperscript{(3)} for more realistic hex GKP synthesis.

\section{Conclusion}

In this work, we have further explored the photon catalysis process~\cite{Lvovsky2002} where interfering a single-photon with an input state on a beamsplitter can result in specifically engineered non-Gaussian quantum states contingent upon beamsplitter parameters and conditioned by PNR detection.  By using a coherent state as an input, we demonstrated that one photon catalysis step can implement arbitrary displacements on the single photon Fock state and preserve state purity.  Further, by cascading the photon catalysis process, we are able to filter out multiple photon number components and accurately approximate relatively large superpositions of squeezed vacuum, including what is to our knowledge the largest proposed optical SCS without the need for large-number Fock states or breeding. Furthermore, we adapted breeding protocols based on homodyne detection~\cite{etesse_proposal_2014,Sychev2017,Lund2004,Oh2018,laghaout_amplification_2013} to a PNR-based scheme that allows for breeding $M$-fold symmetry SSV states. By changing the phase between input modes of the PNR-based breeding, we showed that it is possible to reconstruct square and hexagonal lattice comb-states that exhibit the structure of finite-energy GKP-class code states.  Future avenues of exploration may include utilizing PNR measurement results to implement feed-forward modifications to subsequent photon catalysis steps in the hopes to increase protocol success rates.

It has been recently shown that combining a supply of GKP states with Gaussain QC alone is sufficient for fault-tolerant universal QC; therefore, developing a technique to sythesize GKP states is vital~\cite{baragiola_all-gaussian_2019}. Machine learning optimization algorithms have demonstrated the potential to create high-fidelity GKP states~\cite{Arrazola2019}, but these methods require in excess of 100 gate operations.  The method of photon catalysis and PNR-based breeding we describe here provides an experimentally feasible approach to generate various nonclassical states, including GKP, important for quantum information and quantum computing.

\section*{Acknowledgments}
RN would like to thank Lu Zhang 
for discussions on photon catalysis at QCMC 2018.  ME would like to thank Chun-Hung Chang and Jacob Higgins for pointing out an inconsistency in figure scaling. OP would like to thank Rafael Alexander and Mirko Lobino for discussions. The work is supported by NSF grant No.\ PHY-1708023.

\appendix

\section{Numerical modeling}
\begin{figure}[h]
    \centering
    \includegraphics[width=.6\columnwidth]{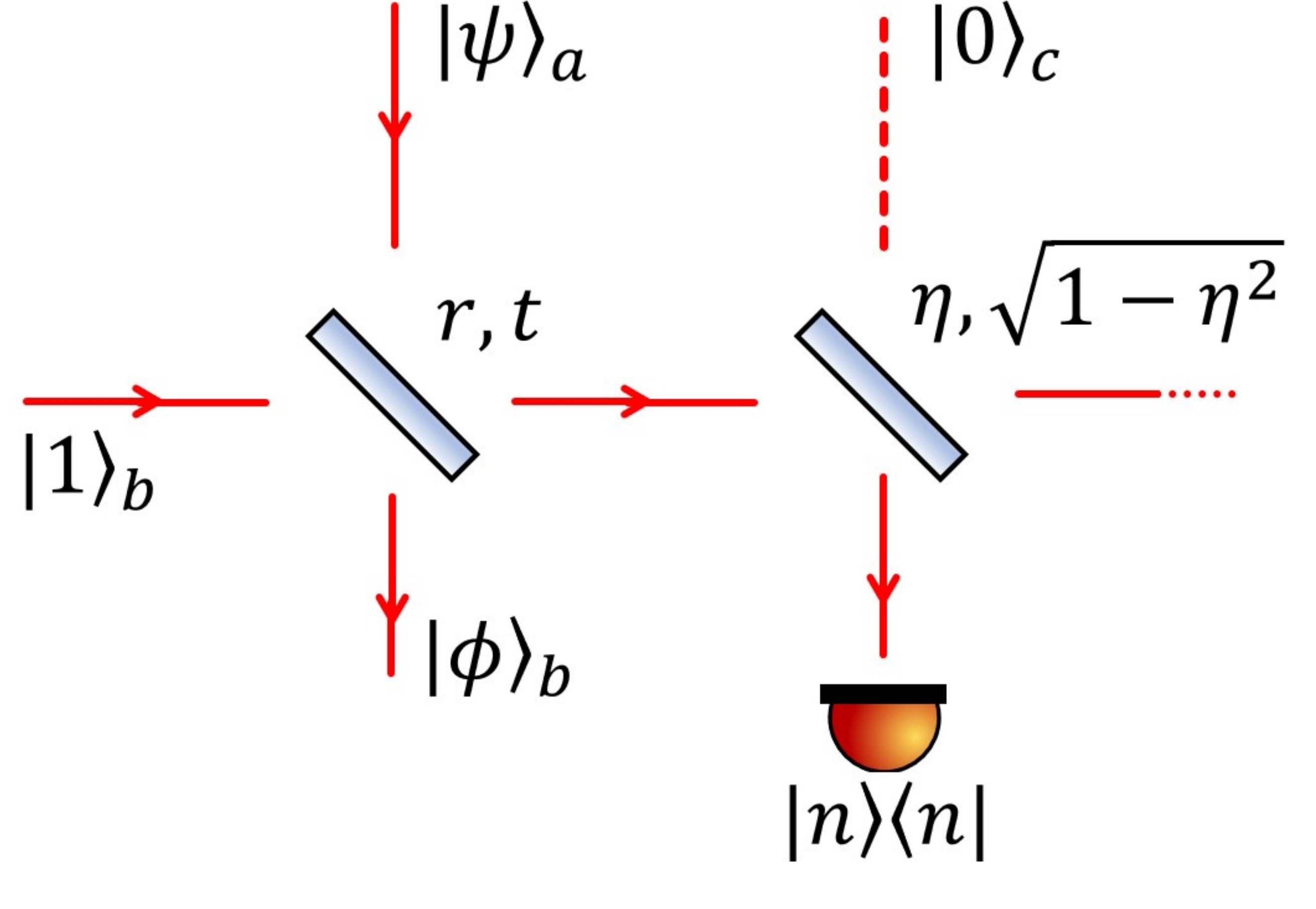}
    \caption{Loss model.  Photon catalysis with an imperfect detector can be modeled by placing a beamsplitter of reflectivity $\eta$ with vacuum input before an ideal detector, where the other output is lost via a partial trace.}
  \label{fig:loss_model} 
\end{figure}

We begin by defining a generic beamsplitter operation, projector, and input density matrices as
\begin{align}
    U_{ab} &= e^{\theta(ab^\dag-a^\dag b)}\\
    P_n &= (|n\rangle\langle n|)_{a}\otimes \mathbb{1}_{b}\\
    \rho^{in}_a &=  (|\psi \rangle\langle \psi|)_a
    \label{eq:rho_in}\\
    \rho^{in}_b &= (|1\rangle\langle1|)_b
\end{align}
where $r=\cos{\theta}$ is the reflection coefficient. After the beamsplitter, but before the detection event, the new density matrix is 
\begin{align}
    \rho_{ab} &= U_{ab}\rho^{in}_a\otimes\rho^{in}_b U_{ab}^\dag.
    \label{eq:rho_new}
\end{align}
We then apply the projective measurement of mode $a$ by an ideal detector and normalize to obtain the resulting density matrix of mode $b$,  given by
\begin{align}
    \rho_{out} &= \frac{Tr_{a}[\rho_{ab}P_n]}{Tr[\rho_{ab}P_n]}
    \label{eq:rho_fin}.
\end{align}
 Imperfect detection can be modeled by including a loss beamsplitter followed by a perfect detector as depicted in Fig.\ref{fig:loss_model}, where we trace out over the lost mode.  In this case we now have a vacuum input mode to consider, so a new input state of $\rho^{in}_{c} = (|0\rangle\langle0|)_c$, a loss beamsplitter $U_{ac} = e^{\xi(ac^\dag-a^\dag c)}$ where $\eta=\cos{\xi}$ is the detector efficiency, and a modified projector, $P^\eta_n = (|n\rangle\langle n|)_{a}\otimes \mathbb{1}_b \otimes \mathbb{1}_c$.   \eq{rho_new} becomes
\begin{align}
    \rho_{abc} &= U_{ac}\rho_{ab}\otimes\rho^{in}_{c} U_{ac}^\dag.
\end{align}
As before, we apply the projective measurement of the detector, but must now also trace out over the lost beamsplitter port in addition to the detected mode.  Our final state is now
\begin{align}
    \rho_{out,\eta} &= \frac{Tr_{ac}[\rho_{abc}P^\eta_n]}{Tr[\rho_{abc}P^\eta_n]}
    \label{eq:rho_loss}.
\end{align}

In several sections of this work, we explore the possibility to perform cascaded photon catalysis by using the resulting density matrix from the $i$\textsuperscript{th} step as the input state to the $i+1$ interference beamsplitter,
\begin{align}
{\rho^{in}_a}^{(i+1)}={\rho^{(i)}_{out}}
\end{align}
This procedure allows us to independently vary the beamsplitter parameter $r_j$ and PNR detection $n_j$ at each step and obtain the intermediate resulting density matrix.  After repeated the process as many times as desired, we can compute the fidelity between a target state, $\rho_T$, and the result of our process, $\rho_{out}$, which is defined by Eq. \ref{eq:fid}.

When performing cascaded photon catalysis calculations, we we limited the Hilbert space dimension to 50 as the tensor product between Hilbert spaces causes an exponential increase in computational expense with each additional mode added.

\section{Cascaded photon catalysis}
Consider iterating the photon catalysis process by using the output from one step as the input to the next.  Here, we derive the general quantum state at the output of an N-step photon catalysis procedure where the initial input is taken to be an arbitrary pure state, $\psi$, and each step uses a single-photon Fock state as the second input. The $i$\textsuperscript{th} step is described by a beamsplitter of real reflectivity and transmissivity, $r_i$ and $t_i$, followed by a perfect PNR detector that detects $n_i$ photons. We start by giving the result and then proceed with mathematical induction as proof. After N steps, we posit that the final quantum state, up to a normalization, will be
\begin{small}
\begin{equation}
\begin{aligned}\label{eq:N_cascaded}
\ket{\phi{_N}} &\propto \sum_{m=0}^\infty\Biggl\{ \psi_{m-N+\sum_{j=1}^{N}n_j}\left[\frac{[m-N+\sum_{j=1}^{N}n_j]!}{m!}\right]^{\frac12}\prod_{k=0}^{N-1}\Biggl[\frac{r_{k+1}^{n_{k+1}-1}}{\sqrt{n_{k+1}!}}t_{N-k}^{m-1+\sum_{j=0}^{k-1}(n_{N-j}-1)}\\
&\times\Bigl(n_{N-k}t_{N-k}^2-r_{N-k}^2(m+\sum_{j=0}^{k-1}(n_{N-j}-1))\Bigr)\Biggr]\Biggr\}\ket m. 
\end{aligned}
\end{equation}
\end{small}
By taking $N=1$, it is easy to see that this reduces to Eq. \ref{eq:2}, thus \ref{eq:N_cascaded} is verified for a single photon catalysis step.  For the proof, we now assume \ref{eq:N_cascaded} to be true and feed $\phi_N$ into an $(N+1)$\textsuperscript{th} step.  The resulting quantum state after the PNR detection of $n_{N+1}$ photons can be obtained from equation \ref{eq:2} by using the $m+n_{N+1}-1$\textsuperscript{th} component of $\phi_N$ to arrive at

\begin{equation}
\begin{aligned}
_{b}\!&\bra{n_{N+1}}U_{ab}\ket{\phi_{N+1}}_a\otimes\ket{1}_b \propto \sum_{m=0}^\infty\Biggl\{ \frac{\psi_{m+n_{N+1}-1-N+\sum_{j=1}^{N}n_j}}{\sqrt{n_{N+1}}}\binom{m+n_{N+1}-1}{m}^{\frac12}\\
&\times\left[\frac{[m+n_{N+1}-1-N+\sum_{j=1}^{N}n_j]!}{(m+n_{N+1}-1)!}\right]^{\frac12}
r_{N+1}^{n_{N+1}-1}t_{N+1}^{m-1}[n_{N+1}t_{N+1}^2-mr_{N+1}^2]\\
&\times\prod_{k=0}^{N-1}\Biggl[\frac{r_{k+1}^{n_{k+1}-1}}{\sqrt{n_{k+1}!}}t_{N-k}^{m+n_{N+1}-2+\sum_{j=0}^{k-1}(n_{N-j}-1)}\\
&\times\Bigl(n_{N-k}t_{N-k}^2-r_{N-k}^2(m+n_{N+1}-1+\sum_{j=0}^{k-1}(n_{N-j}-1))\Bigr)\Biggr]\Biggr\}\ket m. 
\end{aligned}
\end{equation}

Reordering the sum and product indices allows us to arrive at an expression identical to Eq. \ref{eq:N_cascaded}, but with $N\rightarrow N+1$, and thereby complete the proof.  From Eq. \ref{eq:N_cascaded}, we can also extract the overall success probability for a particular set of detection events: 
\begin{equation}
\begin{aligned}\label{eq:success_prob}
&P(n_1,\dots,n_N)= \sum_{m=0}^\infty\Biggl\{ \psi_{m-N+\sum_{j=1}^{N}n_j}^2\left[\frac{[m-N+\sum_{j=1}^{N}n_j]!}{m!}\right]\\
&\times\prod_{k=0}^{N-1}\Biggl[\frac{r_{k+1}^{2n_{k+1}-2}}{n_{k+1}!}t_{N-k}^{2m-2+2\sum_{j=0}^{k-1}(n_{N-j}-1)}
\Bigl(n_{N-k}t_{N-k}^2-r_{N-k}^2(m+\sum_{j=0}^{k-1}(n_{N-j}-1))\Bigr)^2\Biggr]\Biggr\}. 
\end{aligned}
\end{equation}

\section{Fidelity with SSV} \label{sec:AppC}
We now derive an expression for the fidelity between a pure SSV state and the result from an $N$-step filtered coherent state.  The SSV states are given by
\begin{equation}
    \ket{SSV_{\pm}}=C_{SSV\pm}\left[D(\beta)\pm D(-\beta)\right]S(r)\ket0,
\end{equation}
where $r$ is the single-mode squeezing parameter, $\beta$ is the displacement magnitude to create the superposition, and the normalization coefficient is
\begin{equation}
    C_{SSV\pm}=\left[\frac{1}{2\pm2e^{-2|\beta|^2e^{-2r}}}\right]^{\frac12}.
\end{equation}
As a first step, we derive the overlap coefficients between an n-photon Fock state and a displaced m-photon Fock state, which will be useful later.  This is given by
\begin{align}
    A(n,m,\gamma)&\equiv\bra{n}D(\gamma)\ket{m}\\
    &=\bra{n}D(\gamma)\frac{{a^\dag}^m}{\sqrt{m!}}D^\dag(\gamma)\ket{m}\\
    &=\bra{n}\frac{(a^\dag-\gamma^*)^m}{\sqrt{m!}}\ket{\gamma}\\ 
    &=e^{-|\gamma|^2/2}\sum_{k=0}^{m}\binom{m}{k}\frac{(-1)^k\sqrt{n!}\,|\gamma|^{2k+n-m}}{m!(n-m+k)!}.
    \label{eq:overlap}
\end{align}
Because the photon-catalyzed coherent state must be displaced back to the origin of phase space before comparing to an SSV state (see main text), we are interested in a fidelity of the form
\begin{equation}
F=|\bra{SSV_\pm}D^\dag(\delta)\ket{\phi_N}|^2,
\end{equation}
where $\delta$ here is the amplitude of displacement needed after the photon catalysis procedure. By substituting the coefficients of a coherent state of magnitude $\alpha$ for $\psi$ in Eq. \ref{eq:N_cascaded} and using the overlap coefficients given by \ref{eq:overlap}, we can arrive at the overall expression

\begin{equation}
\begin{aligned}
F&=\frac{C_{SSV\pm}^2C_{\ket{\phi_N}}^2}{\cosh{r}}\Biggl|\sum_{\ell,m=0}^\infty\Biggl\{(\tanh{r})^\ell\frac{(2\ell-1)!!\alpha^m}{\sqrt{(2\ell)!m!}}
[A(2\ell,m,-\beta-\delta)-A(2\ell,m,\beta-\delta))]\\
&\times\prod_{k=0}^{N-1}\Bigl[t_{k+1}^m\Bigl(n_{N-k}t_{N-k}^2-r_{N-k}^2(m+\sum_{j=0}^{k-1}(n_{N-j}-1)\Bigr)\Bigr]\Biggr\}\Biggr|^2,
\end{aligned}
\end{equation}
where $C_{\ket{\phi_N}}$ is determined by normalizing Eq. \ref{eq:N_cascaded} with a coherent state input.  This equation for fidelity can now be numerically maximized for the experimentally accessible parameters of $r_i$, $\alpha$, and $\delta$, as well as the desired post-selected PNR measurements of $n_i$ for $N$ photon catalysis steps.

\section{Optimized parameters}\label{sec:data}
The cascaded photon catalysis protocol for $N=2, 3,$ and $4$ steps was numerically optimized to yield the high-fidelity SSV\textsuperscript{(2)} states discussed in the main text. The full list of parameters for the protocols and resultant states are displayed in Table~\ref{tab:label}. 
\begin{table}[H]
\begin{tabularx}{1.0\textwidth}{ |X|X|X|X| } 
  \hline
  $N$ & 2 & 3 & 4 
  \\ 
  \hline 
    Fidelity & $>0.999$ & 0.984 &0.977 
    \\
  \hline 
    Success probability & $1.5\times10^{-2}$ & $1.8\times10^{-3}$ & $4.2\times10^{-5}$
    \\ 
  \hline 
    $\beta$ & 0.90 & 1.35 & 1.59 
    \\
  \hline 
    Squeezing parameter & -0.22 & -0.48 & -0.52   
    \\
  \hline 
    $\alpha$ & 1.20 & 3.54 & 4.66 
    \\
  \hline 
    \{$r_1$,\,$r_2$,\,$r_3$,\,$r_4$\} & \{0.60,\,0.85,\,-\,-,\,-\,-\} & \{0.64,\,0.49,\,0.52,\,-\,-\} & \{0.58,\,0.55,\, 0.70,\,0.42\} 
    \\ 
  \hline 
    \{$n_1$,\,$n_2$,\,$n_3$,\,$n_4$\} &\{1,\,2,\,-\,-,-\,-\}  & \{5,\,2,\,1,\,-\,-\} & \{6,\,4,\,2,\,1\} 
    \\
  \hline 
  \hline
\end{tabularx}
\caption{Parameters of $N$-step photon-catalysis that optimize the fidelity with the nearest SSV\textsuperscript{(2)} state. The fidelity was numerically optimized for all combinations of integer $n_i$ below 10, with all other parameters allowed to vary independently.  The combinations with the highest fidelities and success probabilities are displayed. $\alpha$: input coherent state amplitude. $\beta$: target SSV state amplitude. $r_i$: real-valued reflectiviy of the $i$\textsuperscript{th} beamsplitter in cascaded photon catalysis where $r_i^2+t_i^2=1$. $n_i$: Measurement result of the $i$\textsuperscript{th} PNR detector.}
\label{tab:label}
\end{table}

\section{SSV enlargement by PNR}\label{sec:breeding_appen}
As described in the main text, the breeding protocol for enlarging a general $M$\textsuperscript{th} order symmetry SSV state consists of sending two identical $SSV^{(M)}$ states in respective modes $a$ and $b$ to a balanced beamsplitter, and performing a PNR measurement on the output mode $b$. Before the detection step, the state is given by Eq. \ref{eq:breeding_input} which can be rewritten as
\begin{equation}
\begin{aligned}
\label{eq:E1}
\ket{\phi}&\propto \sum_{n=0}^{M-1}U_{ab}D_a(\beta_n)D_b(\beta_n)S_a(\xi_n)S_b(\xi_n)\ket{0}_a\ket{0}_b\\
&+ \sum_{n_1\neq n_2}^{M-1}U_{ab}D_a(\beta_{n_1})D_b(\beta_{n_2})S_a(\xi_{n_1})S_b(\xi_{n_2})\ket{0}_a\ket{0}_b,\\
\end{aligned}
\end{equation}
where we note that $|\beta_{n_1}|=|\beta_{n_2}|$ and $|\xi_{n_1}|=|\xi_{n_2}|$ for all values of $n_1$ and $n_2$. Using a balanced beamsplitter defined by $U_{ab}=e^{\frac\pi4(ab^\dag-a^\dag b)}$, it can be seen that identical displacers and squeezers on separate modes are transformed in the Heisenberg picture according to
\begin{align}
    U_{ab}D_a(\beta_n)D_b(\beta_n)U^\dag_{ab}&=D_a(\sqrt{2}\beta_n)\\
    U_{ab}S_a(\xi_n)S_b(\xi_n)U^\dag_{ab}&=S_a(\xi_n)S_b(\xi_n),
\end{align}
so after detecting zero photons in mode $b$, the first sum in Eq. \ref{eq:E1} becomes
\begin{equation}
\label{eq:E4}
    \frac{1}{\sqrt{\cosh{|\xi|}}}\sum_{n=0}^{M-1}D_a(\sqrt{2}\beta_n)S_a(\xi_n)\ket{0}_a.
\end{equation}
If we can show that the terms from the second sum in Eq. \ref{eq:E1} are negligible, then normalizing Eq. \ref{eq:E4} will give us the desired enlarged $SSV^{(M)}$ state. The beamsplitter acting on the two unequal displacers gives
\begin{align}
    U_{ab}D_a(\beta_{n_1})D_b(\beta_{n_2})U^\dag_{ab}&=D_a(\frac{\beta_{+}}{\sqrt{2}})D_b(\frac{\beta_{-}}{\sqrt{2}})
\end{align}
where $\beta_+=\beta_{n_1}+\beta_{n_2}$ and $\beta_-=\beta_{n_1}-\beta_{n_2}$. Applying Eq. (13) and (14) from Ref.~\cite{Kim2002} and writing the squeezing arguments as $\xi_{n1}=-re^{2i\varphi}$ and $\xi_{n2}=-re^{2i\theta}$, we see that the two unequal squeezers on vacuum are transformed by
\begin{equation}
\begin{aligned}
    &U_{ab}S_a(-re^{2i\varphi})S_b(-re^{2i\theta})\ket{0}_a\ket{0}_b=R_a(\varphi)R_b(\theta)S_a\Big(-\frac r2 (1+e^{2i(\varphi-\theta)})\Big)\\
    &\times S_b\Big(-\frac r2(1+e^{-2i(\varphi-\theta)})\Big)
    S_{ab}\Big(-2r\sinh{(\varphi-\theta)}\Big)\ket{0}_a\ket{0}_b,
\end{aligned}
\end{equation}

where $R(\Theta)=e^{i\Theta a^\dag a}$ is the phase-shift operator. Detecting zero photon in mode $b$ leads to each term in the second sum of Eq. \ref{eq:E1} taking the form:
\begin{align}
\label{eq:E7}
    D_a(\beta_+)S_a(\varsigma)\,{}_b\!\bra{0}D_b(\beta_-)S_b(\varsigma^*)S_{ab}(\Delta)\ket{0}_b\ket{0}_a, 
\end{align}
where $\varsigma = -\frac r2 (1+e^{2i(\varphi-\theta)}) $ and $\triangle = -2r\sinh{(\varphi-\theta)}. $
Now, realizing that $\beta_-=\frac{|\beta|}{\sqrt{2}}(e^{\frac{2\pi in_1}{M}}-e^{\frac{2\pi i n_2}{M}})$ is always nonzero for $n_1\neq n_2$ when $n_1,n_2<M$, we see that each of the terms having the form of Eq. \ref{eq:E7} consists of the overlap of vacuum with two and single-mode squeezed vacuum that has been displaced.  For finite squeezing, it is clear that this overlap term can always be made arbitrarily small for a large enough displacement amplitude $|\beta_-|$. If this is the case, then the entire second sum in Eq. \ref{eq:E1} vanishes, and detecting zero photons in mode $b$ reduces \ref{eq:E1} to \ref{eq:E4}. 

The only question that remains is how large of an initial displacement amplitude, $|\beta|$, is considered large enough, as it can be seen that increasing $M$ leads to more terms in the sum that we would wish to neglect.  In practice, this can be determined numerically for a desired fidelity threshold with the target enlarged $SSV^{(M)}$, and we see from Fig.\ref{fig:breeding} in the main text that $|\beta|\geq \sqrt{2}$ for $M=2$ and $3$, and $|\beta|\geq \sqrt{5}$ for $M=4$ is sufficient for high-fidelity breeding.

\bibliography{Pfister}

\providecommand{\newblock}{}
\begin{thebibliography}{10}
\expandafter\ifx\csname url\endcsname\relax
  \def\url#1{{\tt #1}}\fi
\expandafter\ifx\csname urlprefix\endcsname\relax\def\urlprefix{URL }\fi
\providecommand{\eprint}[2][]{\url{#2}}

\bibitem{Feynman1982}
Feynman R~P 1982 {\em Int. J. Theor. Phys.\/} {\bf 21} 467--488

\bibitem{Shor1994}
Shor P~W 1994 Algorithms for quantum computation: discrete logarithms and
  factoring {\em Proceedings, $35^{th}$ Annual Symposium on Foundations of
  Computer Science\/} ed Goldwasser S (Santa Fe, NM: IEEE Press, Los Alamitos,
  CA) pp 124--134

\bibitem{Ladd2010}
Ladd T~D, Jelezko F, Laflamme R, Nakamura Y, Monroe C and O'Brien J~L 2010 {\em
  Nature (London)\/} {\bf 464} 45

\bibitem{Lloyd1999}
Lloyd S and Braunstein S~L 1999 {\em Phys. Rev. Lett.\/} {\bf 82} 1784

\bibitem{Braunstein2005a}
Braunstein S~L and van Loock P 2005 {\em Rev. Mod. Phys.\/} {\bf 77} 513

\bibitem{Weedbrook2012}
Weedbrook C, Pirandola S, Garc\'{\i}a-Patr\'on R, Cerf N~J, Ralph T~C, Shapiro
  J~H and Lloyd S 2012 {\em Rev. Mod. Phys.\/} {\bf 84}(2) 621--669
  \urlprefix\url{http://link.aps.org/doi/10.1103/RevModPhys.84.621}

\bibitem{pfister2019continuous}
Pfister O 2019 {\em Journal of Physics B: Atomic, Molecular and Optical
  Physics\/}

\bibitem{Chen2014}
Chen M, Menicucci N~C and Pfister O 2014 {\em Phys. Rev. Lett.\/} {\bf 112}(12)
  120505 \urlprefix\url{http://link.aps.org/doi/10.1103/PhysRevLett.112.120505}

\bibitem{Yoshikawa2016}
Yoshikawa J~i, Yokoyama S, Kaji T, Sornphiphatphong C, Shiozawa Y, Makino K and
  Furusawa A 2016 {\em APL Photonics\/} {\bf 1} 060801
  \urlprefix\url{http://scitation.aip.org/content/aip/journal/app/1/6/10.1063/1.4962732}

\bibitem{Larsen2019}
Larsen M~V, Guo X, Breum C~R, Neergaard-Nielsen J~S and Andersen U~L 2019 {\em
  Science\/} {\bf 366} 369--372 ISSN 0036-8075
  \urlprefix\url{https://science.sciencemag.org/content/366/6463/369}

\bibitem{Asavanant2019}
Asavanant W, Shiozawa Y, Yokoyama S, Charoensombutamon B, Emura H, Alexander
  R~N, Takeda S, Yoshikawa J~i, Menicucci N~C, Yonezawa H {\em et~al.\/} 2019
  {\em Science\/} {\bf 366} 373--376

\bibitem{Menicucci2014ft}
Menicucci N~C 2014 {\em Phys. Rev. Lett.\/} {\bf 112}(12) 120504
  \urlprefix\url{http://link.aps.org/doi/10.1103/PhysRevLett.112.120504}

\bibitem{Gottesman2001}
Gottesman D, Kitaev A and Preskill J 2001 {\em Phys. Rev. A\/} {\bf 64} 012310

\bibitem{Ghose2007}
Ghose S and Sanders B~C 2007 {\em J. Mod. Opt.\/} {\bf 54} 855--869

\bibitem{Bartlett2002}
Bartlett S~D, Sanders B~C, Braunstein S~L and Nemoto K 2002 {\em Phys. Rev.
  Lett.\/} {\bf 88} 097904

\bibitem{Eisert2002}
Eisert J, Scheel S and Plenio M~B 2002 {\em Phys. Rev. Lett.\/} {\bf 89} 137903

\bibitem{Bell1987}
Bell J~S 1987 {\em Speakable and unspeakable in quantum mechanics\/} (Cambridge
  University Press) chap 21, {\em ``EPR correlations and EPW distributions''},
  pp 196--200

\bibitem{Niset2009}
Niset J, {Fiur\'a\v sek} J and Cerf N~J 2009 {\em Phys. Rev. Lett.\/} {\bf 102}
  120501

\bibitem{Vasconcelos2010}
Vasconcelos H~M, Sanz L and Glancy S 2010 {\em Opt. Lett.\/} {\bf 35}
  3261--3263 \urlprefix\url{http://ol.osa.org/abstract.cfm?URI=ol-35-19-3261}

\bibitem{weigand_generating_2018}
Weigand D~J and Terhal B~M 2018 {\em Physical Review A\/} {\bf 97} ISSN
  2469-9926, 2469-9934
  \urlprefix\url{https://link.aps.org/doi/10.1103/PhysRevA.97.022341}

\bibitem{Lvovsky2001}
Lvovsky A~I, Hansen H, Aichele T, Benson O, Mlynek J and Schiller S 2001 {\em
  Phys. Rev. Lett.\/} {\bf 87} 050402

\bibitem{Laiho2010}
Laiho K, Cassemiro K~N, Gross D and Silberhorn C 2010 {\em Phys. Rev. Lett.\/}
  {\bf 105}(25) 253603
  \urlprefix\url{http://link.aps.org/doi/10.1103/PhysRevLett.105.253603}

\bibitem{Morin2012}
Morin O, {D'Auria} V, Fabre C and Laurat J 2012 {\em Opt. Lett.\/} {\bf 37}
  3738--3740 \urlprefix\url{http://ol.osa.org/abstract.cfm?URI=ol-37-17-3738}

\bibitem{Nehra2019}
Nehra R, Win A, Eaton M, Shahrokhshahi R, Sridhar N, Gerrits T, Lita A, Nam S~W
  and Pfister O 2019 {\em Optica\/} {\bf 6} 1356--1360

\bibitem{Dakna1998}
Dakna M, Kn{\"o}ll L and Welsch D~G 1998 {\em Eur. Phys. J. D\/} {\bf 3}
  295--308 ISSN 1434-6060, 1434-6079
  \urlprefix\url{http://www.springerlink.com/openurl.asp?genre=article&id=doi:10.1007/s100530050177}

\bibitem{Ourjoumtsev2006}
Ourjoumtsev A, Tualle-Brouri R, Laurat J and Grangier P 2006 {\em Science\/}
  {\bf 312} 83--86

\bibitem{Zavatta2004}
Zavatta A, Viciani S and Bellini M 2004 {\em Science\/} {\bf 306} 660

\bibitem{Wenger2004}
Wenger J, Tualle-Brouri R and Grangier P 2004 {\em Phys. Rev. Lett.\/} {\bf 92}
  153601

\bibitem{Averchenko2016}
Averchenko V, Jacquard C, Thiel V, Fabre C and Treps N 2016 {\em New J.
  Phys.\/} {\bf 18} 083042
  \urlprefix\url{http://stacks.iop.org/1367-2630/18/i=8/a=083042}

\bibitem{Lvovsky2002}
Lvovsky A~I and Mlynek J 2002 {\em Phys. Rev. Lett.\/} {\bf 88}(25) 250401
  \urlprefix\url{https://link.aps.org/doi/10.1103/PhysRevLett.88.250401}

\bibitem{Pegg1998}
Pegg D~T, Phillips L~S and Barnett S~M 1998 {\em Phys. Rev. Lett.\/} {\bf
  81}(8) 1604--1606
  \urlprefix\url{https://link.aps.org/doi/10.1103/PhysRevLett.81.1604}

\bibitem{bartley_multiphoton_2012}
Bartley T~J, Donati G, Spring J~B, Jin X~M, Barbieri M, Datta A, Smith B~J and
  Walmsley I~A 2012 {\em Physical Review A\/} {\bf 86} ISSN 1050-2947,
  1094-1622 \urlprefix\url{https://link.aps.org/doi/10.1103/PhysRevA.86.043820}

\bibitem{hu_multiphoton_2016}
Hu L~Y, Wu J~N, Liao Z and Zubairy M~S 2016 {\em Journal of Physics B: Atomic,
  Molecular and Optical Physics\/} {\bf 49} 175504 ISSN 0953-4075
  \urlprefix\url{https://doi.org/10.1088%2F0953-4075%2F49%2F17%2F175504}

\bibitem{Birrittella2018}
Birrittella R~J, Baz M~E and Gerry C~C 2018 {\em J. Opt. Soc. Am. B\/} {\bf 35}
  1514--1524
  \urlprefix\url{http://josab.osa.org/abstract.cfm?URI=josab-35-7-1514}

\bibitem{bartley_directly_2015}
Bartley T~J and Walmsley I~A 2015 {\em New Journal of Physics\/} {\bf 17}
  023038 ISSN 1367-2630
  \urlprefix\url{http://stacks.iop.org/1367-2630/17/i=2/a=023038?key=crossref.22d797f0eb412ca30fda3ed5d7615482}

\bibitem{Lita2008}
Lita A~E, Miller A~J and Nam S~W 2008 {\em Opt. Expr.\/} {\bf 16} 3032--3040

\bibitem{Caves1980}
Caves C~M 1980 {\em Phys. Rev. Lett.\/} {\bf 45} 75

\bibitem{BarnettRadmore1997}
Barnett S~M and Radmore P~M 1997 {\em Methods in Theoretical Quantum Optics\/}
  (Oxford: Clarendon Press)

\bibitem{Paris1996}
Paris M~G~A 1996 {\em Phys. Lett. A\/} {\bf 217} 78--80 ISSN 0375-9601
  \urlprefix\url{http://www.sciencedirect.com/science/article/pii/0375960196003398}

\bibitem{kunal2018}
Sharma K and Wilde M~M 2018 Characterizing the performance of
  continuous-variable gaussian quantum gates (\textit{Preprint}
  \eprint{arXiv:1810.12335})

\bibitem{Johansson2012}
Johansson J~R, Nation P~D and Nori F 2012 {\em Comp. Phys. Comm.\/} {\bf 183}
  1760--1772 ISSN 0010-4655
  \urlprefix\url{http://www.sciencedirect.com/science/article/pii/S0010465512000835}

\bibitem{Killoran2019}
Killoran N, Izaac J, Quesada N, Bergholm V, Amy M and Weedbrook C 2019 {\em
  Quantum\/} {\bf 3} 129

\bibitem{Ralph2003}
Ralph T~C, Gilchrist A, Milburn G~J, Munro W~J and Glancy S 2003 {\em Phys.
  Rev. A\/} {\bf 68} 042319

\bibitem{Ourjoumtsev2007}
Ourjoumtsev A, Jeong H, Tualle-Brouri R and Grangier P 2007 {\em Nature
  (London)\/} {\bf 448} 784

\bibitem{Neergaard-Nielsen2006}
Neergaard-Nielsen J~S, Nielsen B~M, Hettich C, M{\o}lmer K and Polzik E~S 2006
  {\em Phys. Rev. Lett.\/} {\bf 97} 083604

\bibitem{Takahashi2008}
Takahashi H, Wakui K, Suzuki S, Takeoka M, Hayasaka K, Furusawa A and Sasaki M
  2008 {\em Phys. Rev. Lett.\/} {\bf 101} 233605
  \urlprefix\url{https://link.aps.org/doi/10.1103/PhysRevLett.101.233605}

\bibitem{Gerrits2010}
Gerrits T, Glancy S, Clement T~S, Calkins B, Lita A~E, Miller A~J, Migdall A~L,
  Nam S~W, Mirin R~P and Knill E 2010 {\em Phys. Rev. A\/} {\bf 82}(3) 031802
  \urlprefix\url{https://link.aps.org/doi/10.1103/PhysRevA.82.031802}

\bibitem{etesse_experimental_2015}
Etesse J, Bouillard M, Kanseri B and Tualle-Brouri R 2015 {\em Physical Review
  Letters\/} {\bf 114} ISSN 0031-9007, 1079-7114
  \urlprefix\url{https://link.aps.org/doi/10.1103/PhysRevLett.114.193602}

\bibitem{Lund2004}
Lund A~P, Jeong H, Ralph T~C and Kim M~S 2004 {\em Phys. Rev. A\/} {\bf 70}
  ISSN 1050-2947, 1094-1622
  \urlprefix\url{https://link.aps.org/doi/10.1103/PhysRevA.70.020101}

\bibitem{Oh2018}
Oh C and Jeong H 2018 {\em J. Opt. Soc. Am. B\/} {\bf 35} 2933--2939
  \urlprefix\url{http://josab.osa.org/abstract.cfm?URI=josab-35-11-2933}

\bibitem{etesse_proposal_2014}
Etesse J, Blandino R, Kanseri B and Tualle-Brouri R 2014 {\em New Journal of
  Physics\/} {\bf 16} 053001 ISSN 1367-2630
  \urlprefix\url{https://doi.org/10.1088%2F1367-2630%2F16%2F5%2F053001}

\bibitem{Sychev2017}
Sychev D~V, Ulanov A~E, Pushkina A~A, Richards M~W, Fedorov I~A and Lvovsky A~I
  2017 {\em Nat. Photon.\/} {\bf 11} 379--382 ISSN 1749-4893
  \urlprefix\url{https://www.nature.com/articles/nphoton.2017.57}

\bibitem{Walls1994}
Walls D~F and Milburn G~J 1994 {\em Quantum Optics\/} (Springer, Berlin)

\bibitem{aharonovich_solid-state_2016}
Aharonovich I, Englund D and Toth M 2016 {\em Nature Photonics\/} {\bf 10}
  631--641 ISSN 1749-4893
  \urlprefix\url{https://www.nature.com/articles/nphoton.2016.186}

\bibitem{loredo2016scalable}
Loredo J~C, Zakaria N~A, Somaschi N, Anton C, De~Santis L, Giesz V, Grange T,
  Broome M~A, Gazzano O, Coppola G {\em et~al.\/} 2016 {\em Optica\/} {\bf 3}
  433--440

\bibitem{somaschi2016near}
Somaschi N, Giesz V, De~Santis L, Loredo J, Almeida M~P, Hornecker G, Portalupi
  S~L, Grange T, Ant{\'o}n C, Demory J {\em et~al.\/} 2016 {\em Nature
  Photonics\/} {\bf 10} 340

\bibitem{ding_-demand_2016}
Ding X, He Y, Duan Z~C, Gregersen N, Chen M~C, Unsleber S, Maier S, Schneider
  C, Kamp M, Höfling S, Lu C~Y and Pan J~W 2016 {\em Physical Review
  Letters\/} {\bf 116} ISSN 0031-9007, 1079-7114
  \urlprefix\url{https://link.aps.org/doi/10.1103/PhysRevLett.116.020401}

\bibitem{heuck_temporally_2018}
Heuck M, Pant M and Englund D~R 2018 {\em New Journal of Physics\/} {\bf 20}
  063046 ISSN 1367-2630
  \urlprefix\url{http://stacks.iop.org/1367-2630/20/i=6/a=063046?key=crossref.fd94a33cb44ec4221511eecbd252efd4}

\bibitem{dauler_review_2014}
Dauler E~A, Grein M~E, Kerman A~J, Marsili F, Miki S, Nam S~W, Shaw M~D, Terai
  H, Verma V~B and Yamashita T 2014 {\em Optical Engineering\/} {\bf 53} 081907
  ISSN 0091-3286, 1560-2303
  \urlprefix\url{https://www.spiedigitallibrary.org/journals/Optical-Engineering/volume-53/issue-8/081907/Review-of-superconducting-nanowire-single-photon-detector-system-design-options/10.1117/1.OE.53.8.081907.short}

\bibitem{Nicolich2019}
Nicolich K~L, Cahall C, Islam N~T, Lafyatis G~P, Kim J, Miller A~J and Gauthier
  D~J 2019 {\em Physical Review Applied\/} {\bf 12} 034020

\bibitem{hanschke2018quantum}
Hanschke L, Fischer K~A, Appel S, Lukin D, Wierzbowski J, Sun S, Trivedi R,
  Vu{\v{c}}kovi{\'c} J, Finley J~J and M{\"u}ller K 2018 {\em npj Quantum
  Information\/} {\bf 4} 43

\bibitem{laghaout_amplification_2013}
Laghaout A, Neergaard-Nielsen J~S, Rigas I, Kragh C, Tipsmark A and Andersen
  U~L 2013 {\em Physical Review A\/} {\bf 87} ISSN 1050-2947, 1094-1622
  \urlprefix\url{https://link.aps.org/doi/10.1103/PhysRevA.87.043826}

\bibitem{Noh2018}
{Noh} K, {Albert} V~V and {Jiang} L 2018 {\em IEEE Trans. Inform. Theory\/}
  1--1 ISSN 0018-9448

\bibitem{glancy2006error}
Glancy S and Knill E 2006 {\em Physical Review A\/} {\bf 73} 012325

\bibitem{albert_performance_2018}
Albert V~V, Noh K, Duivenvoorden K, Young D~J, Brierley R~T, Reinhold P,
  Vuillot C, Li L, Shen C, Girvin S~M, Terhal B~M and Jiang L 2018 {\em
  Physical Review A\/} {\bf 97} 032346
  \urlprefix\url{https://link.aps.org/doi/10.1103/PhysRevA.97.032346}

\bibitem{baragiola_all-gaussian_2019}
Baragiola B~Q, Pantaleoni G, Alexander R~N, Karanjai A and Menicucci N~C 2019
  {\em arXiv:1903.00012 [quant-ph]\/} ArXiv: 1903.00012
  \urlprefix\url{http://arxiv.org/abs/1903.00012}

\bibitem{Arrazola2019}
Arrazola J~M, Bromley T~R, Izaac J, Myers C~R, Brádler K and Killoran N 2019
  {\em Quantum Science and Technology\/} {\bf 4} 024004 ISSN 2058-9565
  \urlprefix\url{http://stacks.iop.org/2058-9565/4/i=2/a=024004?key=crossref.fdbf42ff15681f26130d5962b701f015}

\bibitem{Kim2002}
Kim M, Son W, Bu{\v z}ek V and Knight P 2002 {\em Phys. Rev. A\/} {\bf 65}
  032323

\end{thebibliography}

\end{document}